%% file: ms.tex
\documentclass[twocolumn]{aastex62}

\newcommand{\ChaI}{Cha~I}
\newcommand{\USco}{USco}
\newcommand{\Lup}{Lupus}
\newcommand{\Tau}{Tau/Aur}
\newcommand{\Oph}{Oph}

\newcommand{\ChaIFull}{Chamaeleon~I}
\newcommand{\UScoFull}{Upper Scorpius OB}
\newcommand{\LupFull}{Lupus}
\newcommand{\TauFull}{Taurus \& Auriga Complex}
\newcommand{\OphFull}{Ophiuchus}

\newcommand{\Lmm}{\ensuremath{%
    L_{\text{mm}}}%
}

\newcommand{\Mstar}{\ensuremath{%
    M_{\star}}%
}

\newcommand{\Mdust}{\ensuremath{%
    M_{\text{dust}}}%
}

\newcommand{\Lstar}{\ensuremath{%
    L_{\star}}%
}

\newcommand{\R}[1]{\ensuremath{%
    R_{\text{#1}}}%
}

\newcommand{\RsixvsLstar}{\ensuremath{%
    \log{R_{68}}-\log{\Lstar}}%
}

\newcommand{\RsixvsMstar}{\ensuremath{%
    \log{R_{68}}-\log{\Mstar}}%
}

\newcommand{\RsixvsLmm}{\ensuremath{%
    \log{R_{68}}-\log{\Lmm}}%
}

\newcommand{\RvsLmm}{\ensuremath{%
    \log{R_{\text{eff}}}-\log{\Lmm}}%
}

\newcommand{\RvsMstar}{\ensuremath{%
    \log{R_{\text{eff}}}-\log{\Mstar}}%
}

\newcommand{\RvsLstar}{\ensuremath{%
    \log{R_{\text{eff}}}-\log{\Lstar}}%
}

\newcommand{\LmmvsLstar}{\ensuremath{%
    \log{\Lmm{}}-\log{\Lstar{}}}%
}
\newcommand{\LmmvsMstar}{\ensuremath{%
    \log{\Lmm{}}-\log{\Mstar{}}}%
}
\newcommand{\LstarvsLmm}{\ensuremath{%
    \log{\Lmm{}}-\log{\Lstar{}}}%
}

\newcommand{\MdustvsMstar}{\ensuremath{%
    \log{M_{\text{dust}}}-\log{\Mstar{}}}%
}

\newcommand{\RpropL}[1]{\ensuremath{%
    \R{eff} \propto \Lmm^{#1}}
}

\newcommand{\dem}[1]{\textcolor{gray}{#1}}

\newcommand{\Lsun}{L_{\odot}}
 
\graphicspath{{./}{figures/}}

\usepackage[section]{placeins}

\begin{document}

\title{The evolution of dust-disk sizes from a homogeneous analysis of 1-10 Myr-old stars}

\correspondingauthor{Nathanial Hendler}
\email{nathanhendler@gmail.com}

\author[0000-0002-3164-0428]{Nathanial Hendler}
\affil{Lunar and Planetary Laboratory, The University of Arizona, Tucson, AZ 85721, USA}

\author[0000-0001-7962-1683]{Ilaria Pascucci}
\affiliation{Lunar and Planetary Laboratory, The University of Arizona, Tucson, AZ 85721, USA}

\author[0000-0001-8764-1780]{Paola Pinilla}
\affiliation{Max-Planck-Institut f\"{u}r Astronomie, K\"{o}nigstuhl 17, 69117, Heidelberg, Germany}

\author[0000-0003-3590-5814]{Marco Tazzari}
\affiliation{Institute of Astronomy, University of Cambridge, Madingley Road, CB3 0HA Cambridge, UK}

\author[0000-0003-2251-0602]{John Carpenter}
\affiliation{Joint ALMA Observatory, Avenida Alonso de C\'ordova 3107, Vitacura, Santiago, Chile}

\author[0000-0002-1226-3305]{Renu Malhotra}
\affiliation{Lunar and Planetary Laboratory, The University of Arizona, Tucson, AZ 85721, USA}

\author[0000-0003-1859-3070]{Leonardo Testi}
\affiliation{European Southern Observatory, Karl-Schwarzschild-Strasse 2, 85748 Garching bei München, Germany}



\begin{abstract}

    We utilize ALMA archival data to estimate the dust disk size of 152
    protoplanetary disks in \Lup{} (1-3\,Myr), Chamaeleon I (2-3\,Myr), and
    Upper-Sco (5-11 Myr).  We combine our sample with 47 disks from \Tau{} and
    \Oph{} whose dust disk radii were estimated, as here, through fitting
    radial profile models to visibility data.  We use these 199 homogeneously
    derived disk sizes to identify empirical disk-disk and disk-host property
    relations as well as to search for evolutionary trends.  In agreement with
    previous studies, we find that dust disk sizes and millimeter luminosities
    are correlated, but show for the first time that the relationship is not
    universal between regions.  We find that disks in the 2-3\,Myr-old \ChaI{}
    are not smaller than disks in other regions of  similar age, and confirm
    the \cite{Barenfeld2017} finding that the 5-10\,Myr \USco{} disks are
    smaller than disks belonging to younger regions.  Finally, we find that the
    outer edge of the Solar System, as defined by the Kuiper Belt, is
    consistent with a population of dust disk sizes which have not experienced
    significant truncation.

\end{abstract}




\section{Introduction} \label{sec_intro}

Protoplanetary disks, consisting of gas and dust around young ($\sim$1--10~Myr)
stars, are the sites of planet formation.  Because the expected population of
planetesimals (km-sized bodies) or larger planetary embryos 
within these disks is not directly observable, we rely on millimeter
observations sensitive to the largest detectable dust grains to constrain the
timing, location, and mechanics of planet formation.  The radial distribution
of these mm/cm sized grains within the disk is a key parameter governing the
planet making potential of a disk.  For instance, in the pebble accretion scenario, the total amount of millimeter-sized
grains and their inward flux are critical to form planets \citep[e.g][]{Ormel2017}. 

Millimeter surveys of protoplanetary disks reveal typical disk properties, as
well as  their spread, and can be used to identify empirical relationships
between disk properties and disk/host-star properties.  Such relations are
essential to test and inform planet formation
\citep[e.g.][]{Mulders2015,Pascucci2018}, as well as in understanding the
diversity of observed exo-planetary systems.

ALMA surveys of the nearby low-mass star-forming regions of \Lup{}, \ChaI{}
and \USco{} \citep[][respectively]{Ansdell2016,Pascucci2016,Barenfeld2016} each
found a positive correlation between the mass in millimeter grains
(hereafter,\Mdust{}) and stellar mass (hereafter, \Mstar{}) for their
respective region.  It was also shown that the relationship steepens with the
age of the region \citep{Pascucci2016}, suggesting that the amount of pebbles
available to form planets decreases faster for disks around low-mass stars.

Pre-ALMA observations of the brightest disks in different star-forming regions
reported dust disk outer radii ranging from 22 to 440~au
\citep{Isella2009,Andrews2010,Guilloteau2011}.  In addition, a correlation
between disk size and disk luminosity (\Lmm{}) was identified early on
\citep{Andrews2010}.
However, only the $\sim3$ times larger SMA sample analyzed by
\cite{Tripathi2017} could quantify the relation and found that \Lmm{} scales as
the square of the dust disk radius.

Dust disk size estimates from ALMA surveys have been carried out for  \Lup{}
\citep{Tazzari2017, Andrews2018a}, \USco{} \citep{Barenfeld2017} and the Orion
Nebula Cluster \citep{Eisner2018}, yet each has been performed with 
different
modeling techniques and assumptions.  \cite{Tazzari2017} estimated \Lup{} disk
sizes by fitting a two layer model \citep{Chiang1997} to visibilities, and
found that \Lup{} disks tend to be larger than previously reported disk sizes in
\Tau{} and \Oph{} observed at similar angular resolutions.
\cite{Andrews2018a} estimated the disk sizes of \Lup{}
by fitting Nuker brightness profiles, and compared them to the disk sizes of \Tau{} and
\Oph{} as estimated by \cite{Tripathi2017} and did not come to the same
conclusion that the \Lup{} disks are generally larger.  While both works
analyzed sub-samples of the same ALMA campaign, the \cite{Andrews2018a} sample
was $\sim2$\, times larger than the \cite{Tazzari2017} sample which had a 4\,mJy cutoff, excluded
unresolved disks as well as disks with resolved gaps or cavities.  However,
both works determined that disks within \Lup{} have a positive correlation
between disk size and \Lmm{}.  \cite{Barenfeld2017} estimated the sizes of
disks within the older \USco{} region by fitting radiative transfer models to visibilities
using a truncated power law for surface density.  They found
\USco{} disks to be typically three times smaller than the disks in \Oph{},
\Tau{} and \Lup{}, and suggested that \USco{} followed the same positive correlation
between disk size and \Lmm{} as \cite{Tripathi2017}.  Finally, \cite{Eisner2018} measured disks within the
Orion Nebula Cluster by fitting a 2D elliptical Gaussian to ALMA continuum
maps.  Their work found a correlation between disk size and \Lmm{} as well, and
additionally suggested that \ChaI{} disks, using HWHM measurements from
\cite{Pascucci2016} for \ChaI{} disk sizes, are significantly smaller than
\Oph{}, \Tau{}, and \Lup{}.

While trends seen within each region are robust, inferences among regions 
might be compromised by the use of
varying modeling techniques (i.e. image plane measurements, radiative transfer
modeling, visibility fitting with profiles generated by different functions)
and assumptions (i.e. opacity, paramaterized disk height, fixed surface density
slope, disk temperature, inner radius location).  \cite{Andrews2018a} addressed
this problem by comparing the regions of \Oph{}, \Tau{} and \Lup{} in a
homogeneous way, finding that mm-luminosity scales as the square of the
dust-disk radius was common to all three regions.  This suggests that the
scaling law between disk radius and stellar luminosity may be universal.

In this work we expand on the results of \cite{Tripathi2017} and
\cite{Andrews2018a}, using the same modeling techniques in order to present a
homogeneously derived census of disk sizes within five regions of varying ages.
In Section \ref{sec_sample} we discuss our sample selection, consisting of a
combination of disk size estimates from literature for \Oph{} and \Tau{}
(Section \ref{sec_sample_lit}) with our own estimates for disks within \Lup{},
\ChaI{} and \USco{} (Section \ref{sec_sample_alma}).  Our reduction of the ALMA observations is discussed in Section \ref{sec_observations}.  To estimate disk sizes,
we model observed sky brightnesses using axisymmetric radial-profile models
which are fit to ALMA visibilities (see Section \ref{sec_modeling}).
Because not every source is detected, or results in a model that provides a
disk size estimate, Section \ref{sec_modeling_sample_selection} provides
details on the selection criteria we use for the final sample of disks used in
our analysis.
In Section \ref{sec_results} we summarize our results and
compare disk sizes between regions.  We use a variety of statistical tests
to assess if relationships exist
between disk properties and stellar-host properties in Section \ref{sec_relations}.
Finally, we discuss and summarize our results in Sections \ref{sec_discussion} and \ref{sec_summary}.

\section{Initial Sample Selection} \label{sec_sample} We aim to obtain a
representative sample of stars with protoplanetary disks from different
regions spanning a range of ages. In order to make a
proper comparison, we have also chosen regions observed at similar wavelengths
and spatial scales (see Table~\ref{tab_sample}).  Each of the disks included in
this work belongs to the following 5 regions: \OphFull{} which is
$\sim$\,1-2\,Myr \citep{Wilking2005, Luhman1999}, hereafter Oph; the \TauFull{}
which is $\sim$\,1-3\,Myr \citep{Luhman2004}, hereafter Tau/Aur; \LupFull{} which
is $\sim$\,1-3\,Myr \citep{Comeron2008,Alcala2014}; \ChaIFull{}  which is
$\sim$\,2-3\,Myr \citep{Luhman2008}, hereafter Cha~I; and the \UScoFull{}
association  which is $\sim$\,5-11\,Myr \citep{Preibisch2002, Pecaut2012,
Slesnick2008}, hereafter USco.  

We use previously published disk sizes for Oph and Tau/Aur with observations
obtained by the Submillimeter Array (SMA; see
Section~\ref{sec_sample_lit} for more details), while we estimate sizes for
disks in Lupus, Cha~I, and USco from archival ALMA data
(Section~\ref{sec_sample_alma}). Stellar masses are also derived homogeneously
for the latter three regions after re-scaling literature stellar luminosities
to the Gaia DR2 distances (Section~\ref{sec_stellar_properties}).

\begin{deluxetable}{llcccr}
\tablecaption{Regions included in our analysis\label{tab_sample}}
\tablewidth{0pt}
\tablehead{
    \colhead{Region}     & 
    \colhead{Telescope}  & 
    \colhead{$\lambda$}  & 
    \colhead{Typical Beam} \\
    \colhead{}          & 
    \colhead{}          & 
    \colhead{(\micron)}   & 
    \colhead{(arcsec)}    & 
}
\startdata
    \Oph{}  & SMA  & 880     & $0.41 - 0.78 $ \\
    \Tau{}  & SMA  & 880     & $0.41 - 0.78 $ \\
    \Lup{}  & ALMA & 935-954 & $0.28 - 0.35$  \\
    \ChaI{} & ALMA & 884-887 & $0.5 - 0.7$    \\
    \USco{} & ALMA & 876-975 & $0.35 - 0.75$  \\
\enddata
\end{deluxetable}

\subsection{\Oph{} and \Tau{}} \label{sec_sample_lit}

We include in our analysis previously derived disk sizes for sources in the
\Oph{} and \Tau{} star-forming regions \citep{Tripathi2017}, relying on the
Gaia-updated values reported in \citet{Andrews2018a}.  From the entire sample
of 50 disks we exclude TW Hya, HD 163296, and LkH$\alpha$~330 because they do
not belong to \Oph{} nor \Tau{}.

\cite{Andrews2018a} derived disk sizes also for 56 sources belonging to
the \Lup{} star-forming region, from ALMA observations originally presented in
\citet{Ansdell2016}.  We use these literature values only to verify that our
modeling approach delivers the same results (Section~\ref{sec_sample_alma} and
Appendix \ref{app_lupus_comparison}) and justify the extended comparison of
disk sizes carried out here. We note, however, that
the \Oph{} and \Tau{} samples come from flux-limited SMA observations, and
consequently are biased towards brighter objects than the \Lup{}, \ChaI{}, and
\USco{} samples.  We take this bias into account when interpreting the results.

\subsection{\Lup{}, \ChaI{}, and \USco{}} \label{sec_sample_alma}

To test if our approach for estimating disk sizes delivers the same results as
\cite{Tripathi2017} and \cite{Andrews2018a}, we re-reduce and re-analyze the
ALMA data from the Lupus star-forming region.  We include in our sample all the
62 detections from the ALMA project \texttt{2013.1.00220.S} (PI: Johnathan
Williams) as presented in \cite{Ansdell2016}.
As discussed
in Appendix \ref{app_lupus_comparison} and shown in
Figures~\ref{fig_andrews_vs_me} and \ref{fig_Lmm_andrews_vs_me}, our method
retrieves the same disk sizes as those in \cite{Andrews2018a} for most sources,
as well as the same disk size-millimeter luminosity relation within the quoted
uncertainties.  Because the uncertainties reported in \cite{Andrews2018a} are
systematically lower than ours (see Appendix \ref{app_lupus_comparison}), we
prefer to use the inferred disk sizes from our modeling in the analysis and
discussion sections of this paper in order to have consistent uncertainties
across all of our samples.

For the \ChaI{} region, we include as part of our sample the 66 detections
reported in \cite{Pascucci2016} from ALMA project \texttt{2013.1.00437.S} (PI:
Pascucci, I.).  Five detections and nine non-detections from that project were
re-observed at 5 times higher sensitivity in project \texttt{2015.1.00333.S}
(PI: Pascucci, I.).  We use the 11 detections reported in \cite{Long2018},
giving us a total of 72 \ChaI{} disks.

Finally, \USco{}  was observed in projects \texttt{2011.0.00526.S} and
\texttt{2013.1.00395.S} (PI: Carpenter, J.) and results were originally
presented in \cite{Carpenter2014} and \cite{Barenfeld2016} respectively.  To
estimate disk sizes we have selected those 50 USco targets that were detected.

\subsubsection{Stellar masses} \label{sec_stellar_properties}

For \Lup{}, \ChaI{}, and \USco{} we also re-derive stellar masses in order to
(a) have self-consistent values for the ALMA datasets, and (b) take advantage
of the newer Gaia DR2 distances.  Stellar masses are determined following the
Bayesian inference approach described in \cite{Pascucci2016}.  First, we
collect stellar effective temperatures and bolometric luminosities from the
literature: For \Lup{} we rely on
\cite{Alcala2014,Biazzo2017,Frasca2017,Andrews2018a}; For \ChaI{} on
\cite{Manara2016,Manara2017}; while for \USco{} we rely on
\cite{Barenfeld2016}.  Then, we scale these luminosities to the new Gaia DR2
distances.  For sources in \ChaI{} and \USco{} we query distances from the
\cite{Bailer-Jones2018} Gaia catalogue.  When there is no DR2 distance
available, we use the median sample distance of 190\,pc for  \ChaI{}  and
144\,pc for \USco{}, both of which agree with the values obtained from all
members of each region (see \citealt{Roccatagliata2018} and
\citealt{deZeeuw1999}).  For our \Lup{} sample, we take the GAIA DR2 distances
as presented in \cite{Andrews2018a}.  J11072825-7652118 (\ChaI{}) and
J16141107-2305362 (\USco{}) have anomalously large DR2 distances (744\,pc, and
6011\,pc respectively).  For these three sources we also use the median
distance of each region as given above.  Following \cite{Pascucci2016}, we
assume an uncertainty of 0.02\,dex in the stellar temperature for spectral
types earlier than M3 and 0.01\,dex for later spectral types and a 0.1\,dex
uncertainty on all stellar luminosities.  Table \ref{tab_stellar} within
Appendix \ref{app_properties} summarizes the adopted and inferred
stellar parameters for \Lup{}, \ChaI{}, and \USco .

\section{ALMA Observations and Data Reduction: Lupus, Cha I and USco} \label{sec_observations}

We re-reduce band 7 ($\sim880-975\micron$) observations of similar sensitivity
for the sample of disks presented in Section~\ref{sec_sample_alma}.  The data
reduction steps that lead to the calibrated visibilities are described below.

In general, the raw data is taken from the ALMA archive and measurement sets
are built using the calibration scripts created by the North American ALMA
Science Center (NAASC).  For this step we use the same version of the Common
Astronomy Software Applications (CASA, \citealt{McMullin2007}) as used by the
NAASC noted within the downloaded scripts.  These scripts perform phase,
bandpass and flux calibrations and create the standard CASA measurement sets
containing the visibilities.

Because the 8 outermost edge channels in each spectral window are typically
noisy, we remove them from our measurement sets.  Next, we average the
measurement sets in time and by channel.  The parameters used to do time
averaging and spectral window averaging are unique to each source with the goal
of ending up with similarly sized (in number of data points) data sets.
Typically, we average spectral windows by widths of 19 channels.  We average
our data in time by a variable number of seconds based on the total amount of
data available to us (this varies by exposure time and number of baselines).
However, we constrain all time averaging to be between 2 and 30 seconds.  For
these steps, we use \texttt{CASA} version \texttt{4.7.2-el6}.

Measurement sets from the project \texttt{2011.0.00526.S} (\USco{}) were
directly provided by the PI and co-author J. Carpenter.  For these sources we
used a width of 22 channels for the spectral window averaging.

Calibration of the sources in projects \texttt{2015.1.00333.S} (\ChaI{}) and
\texttt{2013.1.00395.S} (\USco{}) were performed using \texttt{CASA} version
\texttt{4.7.2-el6}.  Calibration of the sources in science goal
\texttt{A001\_X11d\_X13} (\ChaI{}) and project \texttt{2013.1.00220.S} (\Lup{})
was performed using \texttt{CASA} version \texttt{5.1.1-5.el7}.  These
exceptions were required due to incompatibilities between the NAASC provided
calibration scripts and system libraries.

In addition, for the 10 brightest \ChaI{} disks\footnote{These disks are:
J10581677--7717170, J10590699--7701404, J11022491--7733357, J11040909--7627193,
J11074366--7739411, J11080297--7738425, J11081509--7733531, J11092379--7623207,
J11094742--7726290, and J11100010--7634578.}, we use self-calibrated visibilities
from \cite{Pascucci2016}.  The disk sizes resulting from self-calibrated
visibilities are the same as those derived from non-self-calibrated
visibilities, hence we do not apply self calibration to the other regions.

\section{Modeling Method} \label{sec_modeling}

In this section we describe our approach to model the calibrated continuum visibilities
and our procedure for determining the dust disk outer radius (\R{eff}).
To expand upon the results presented in \cite{Tripathi2017} and \cite{Andrews2018a}, we use a similar method to derive disk sizes for Lupus, ChaI and USco.

As high-contrast asymmetries seem to be uncommon even in high-resolution ALMA images
\citep{Long2018,Andrews2018}, we assume axisymmetric disks and use a parametric
radial intensity profile to fit the
disk intensity
in the visibility domain.

For all modeling, we fit disk inclination ($i$), position angle (PA), and disk
center offsets (dRA, dDec) in addition to the free parameters connected to the
chosen radial profile function described  below.  All modeling also includes 
a \textit{nuisance} parameter (lnwcorr) defined as the natural logarithm of the factor by which the weights of the observed visibilities are overestimated. 

For all disks, we test two different functions for generating the radial
profile shape: a Nuker profile \citep{Lauer1995} and a Dirac delta function (point
source).  By comparing the best-fit models of each function (using the reduced $\chi^2$) for a given source,
we are able to determine if the disk is resolved (Nuker fits better) or not
(point source fits better). 

The Nuker profile was shown by \cite{Tripathi2017} to be a useful radial
profile function thanks to its ability to reproduce the
sky brightness of both full and transition disks.   The radial intensity of a Nuker profile is expressed as:

\begin{equation} \label{eq_radial_profile}
    I(r) = F_0\bigg(\frac{r}{\R{t}}\bigg)^{-\gamma}\bigg[1 + \bigg(\frac{r}{\R{t}}\bigg)^{\alpha}\bigg]^{(\gamma-\beta)/\alpha}
\end{equation}

where $r$ is the radial distance, $F_0$ is the amplitude coefficient, $\R{t}$
is the transition radius, $\alpha$ is the transition index, while  $\gamma$ and
$\beta$ are indexes that define the inner and outer cutoff, respectively. 
Figure~2 in \cite{Tripathi2017} nicely illustrates how each parameter affects
the shape of the Nuker profile.

\begin{figure*}
    \plotone{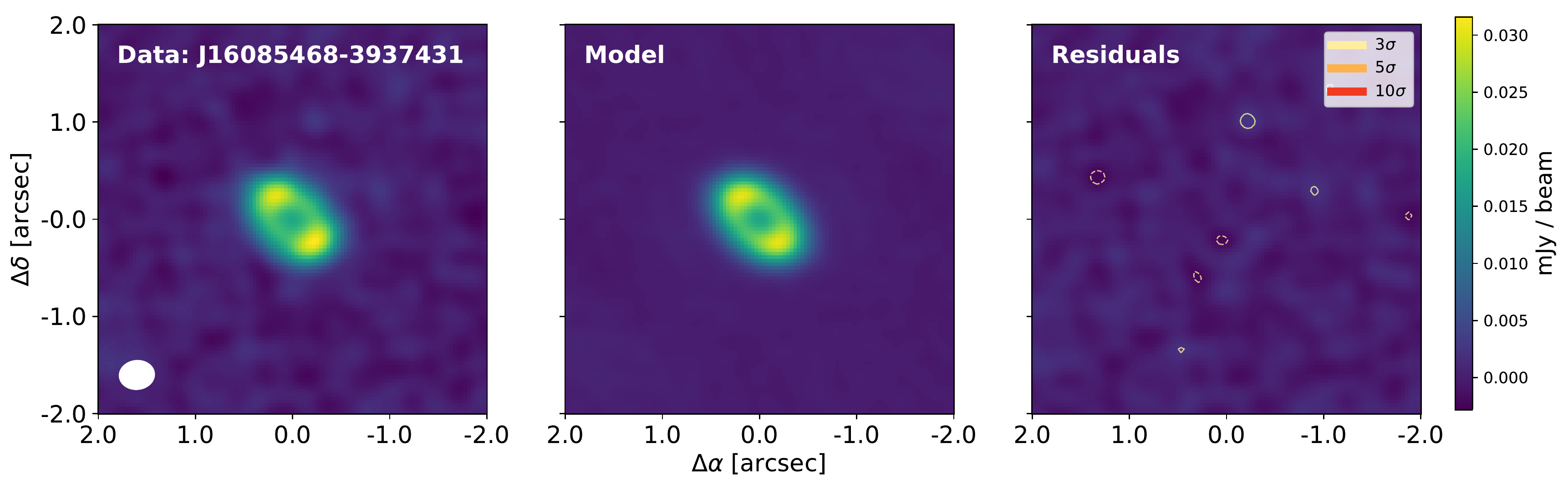}
    \caption{Comparison of the ALMA $887~\micron$ observation of J16085468-3937431 with our best-fit model.  The first panel shows the ALMA observation as a continuum map generated using the CASA \texttt{clean} command with Briggs weighting with a robustness parameter of 0.5.  The middle panel shows a continuum map generated from our model using the same UV spacings as the ALMA observation.  The last panel shows the residuals between the first two panels (residuals of 3, 5, and 10 sigma are denoted as outlines).
    \label{fig_residuals}}
\end{figure*}

\begin{figure}
    \plotone{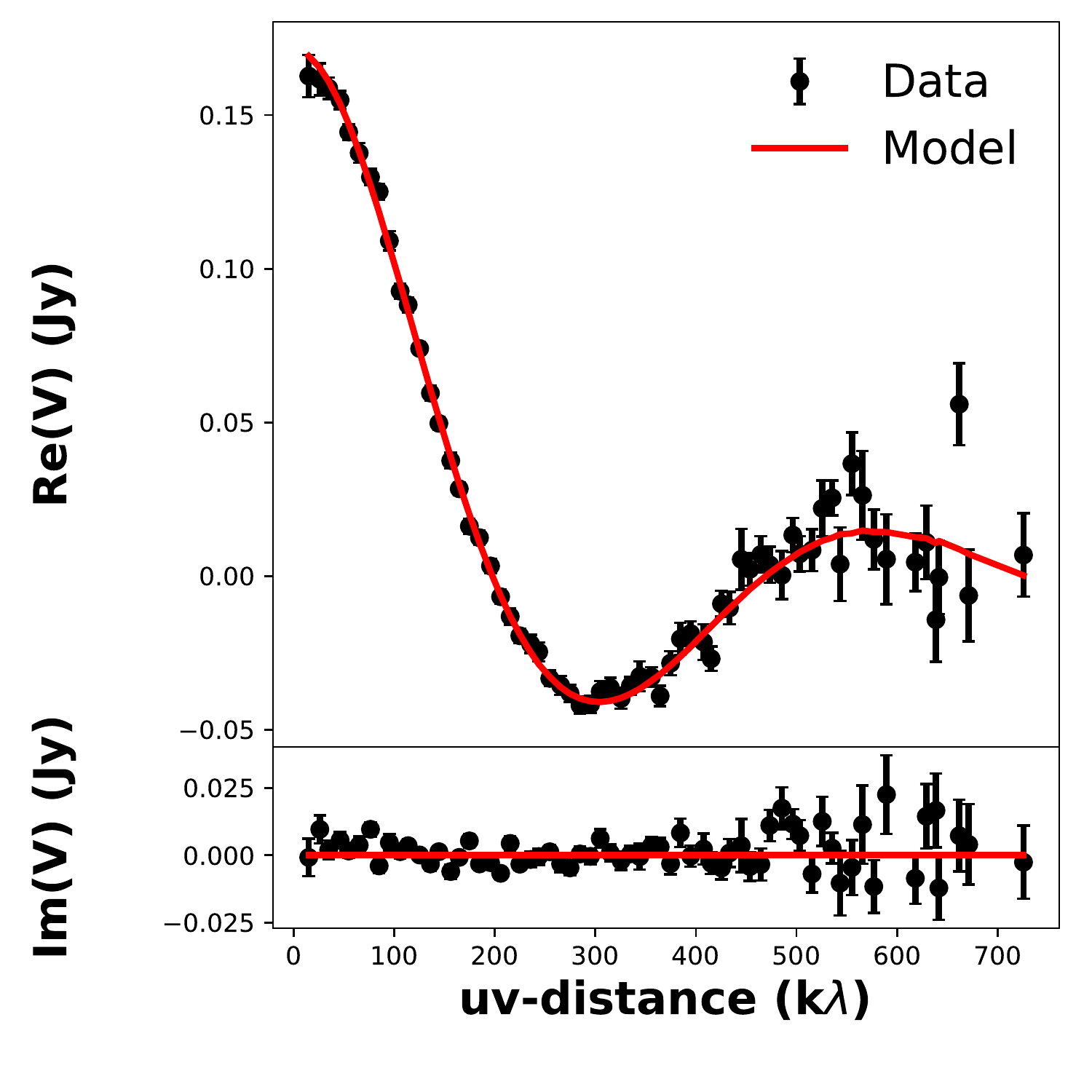}
    \caption{A comparison of our model with the observed visibilities for the source J16085468-3937431.  The real and imaginary components of the observed visibilities (filled circles) are azimuthally averaged and deprojected. For the clarity of the figure, the visibility data is further binned in increments of $10\text{k}\lambda$. Our best-fit (Nuker profile) model is plotted in red. \label{fig_uvfitting} Plot made with the \textsc{uvplot} package \citep{Tazzariuvplot}.}
\end{figure}

For a given profile (Nuker or point source), we produce a synthetic disk image
which is Fourier transformed and sampled at the same spatial frequencies as our
observational data using \texttt{GALARIO} \citep{Tazzari2018}.  Our modeled
visibilities are fit to the real and imaginary parts of the observed
visibilities using the \texttt{emcee} implementation \citep{Foreman-Mackey2013} 
of the Markov Chain Monte Carlo method (MCMC).
Uniform priors are used over a parameter space defined as:
$\R{t} \in [0.005,3]$\,au,
$\gamma \in [-11,4]$,
$\log{\alpha} [0.3,1.3]$,
$\beta \in [1,17]$,
$\log{F_0}  \in [3,12]$\,Jy/Sr, 
$i$        $\in [0,90]$\,deg, 
PA         $\in [0,179]$\,deg, 
dRA        $\in [-3,3]$\,arcsec, 
dDec       $\in [-3,3]$\,arcsec, 
lnwcorr     $\in [-10,10]$.
We cover this parameter space
with 70 chains (ensemble sample walkers) which individually take 100,000 steps
in order to sample the posterior probability distribution function (PDF).  The
location of each walker is initialized using random draws from a truncated
normal distribution about the median value for each parameters following an
initial MCMC burn in cycle using 100 walkers and 1000 steps.  An example best
fit model for the source J16085468-3937431, is given in Figures
\ref{fig_residuals} and~\ref{fig_uvfitting}, all other fits are provided
in the electronic version of this paper.

\begin{figure*}[ht]
    \plotone{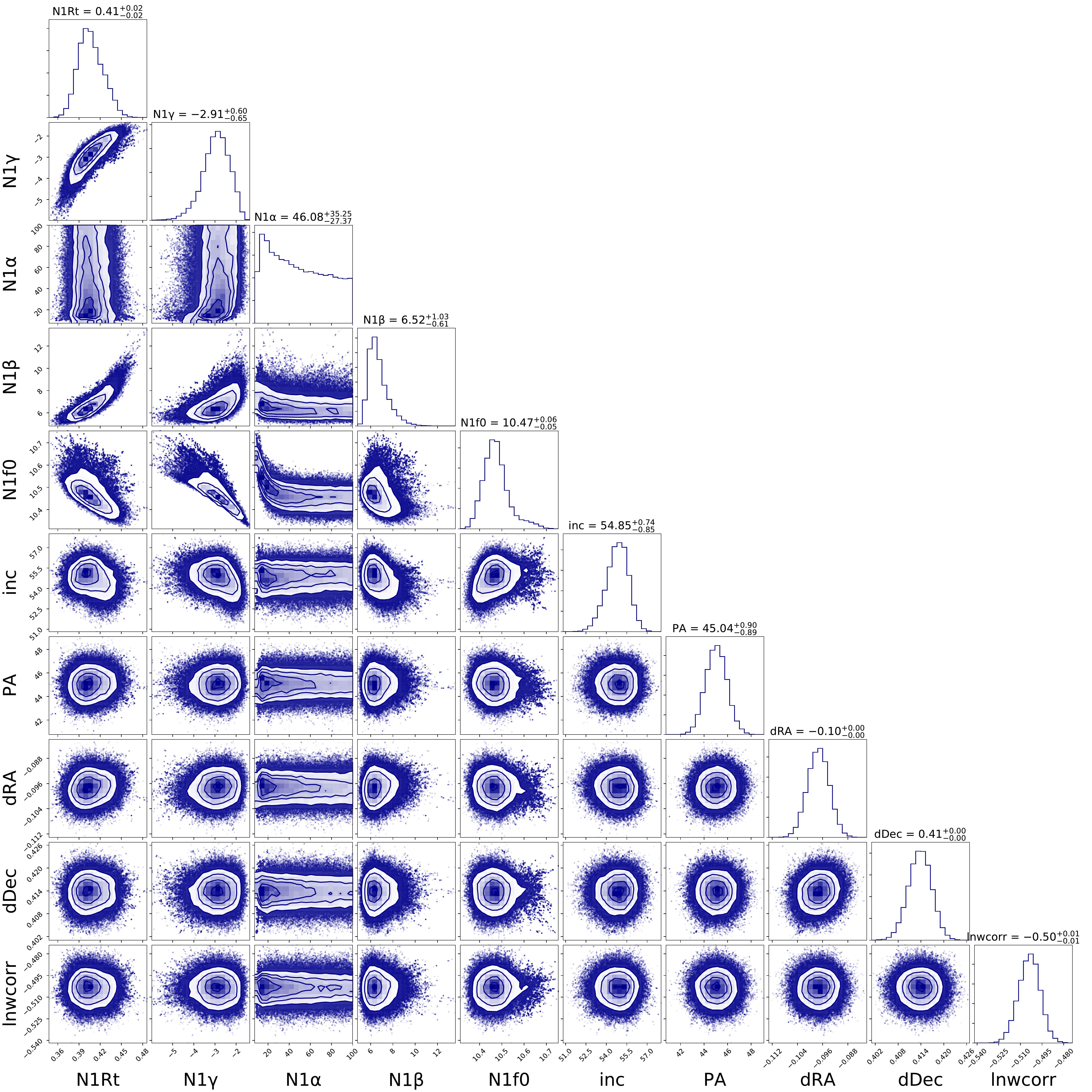}
    \caption{
        Corner plot of the free parameters generated from the MCMC fitting of J16085468-3937431.  Marginalized distributions are shown as histograms at the top of each column.  Parameters with names preceded by ``N1'' refer to the corresponding Nuker parameters (see Equation \ref{eq_radial_profile}).  Parameters used to deproject and center the disk are inclination (inc), position angle (PA), right ascension offset (dRA), and declination offset (dDec).  The last column is the fitted weight correction factor (lnwcorr; see Section \ref{sec_modeling}).
        \label{fig_corner}}
\end{figure*}

MCMC fitting produces a chain of models.  We use the autocorrelation length as
a guide to understand at what point in the chain convergence occurs.  This
typically happens well before $10^4$ steps per walker, leaving us with
approximately $6.99\times10^6$ samples per disk.  We then conservatively ignore
half of the converged chain length and only consider the $3.5\times10^6$
samples from the end of the MCMC chain to estimate the posterior probability
density function (PDF), and parameter uncertainties.

The bounds of the parameter space explored by the MCMC walkers for the
variables $i$, PA, dRA, dDec and lnwcorr is the same for all regions.  The
remaining free parameters explored is determined by the radial profile being
modeld (Nuker or point-source).  For these parameters, the bounds are adjusted
for each star-forming region based on prior exploration of the parameter space
to ensure no truncation of the posterior distributions.  The boundaries we
chose are similar to those in \cite{Tripathi2017} and
\cite{Andrews2018}, and are chosen to cover well the posterior distribution.

At this point it is important to point out that the physical parameter we are
interested in, disk size, is not one of the free parameters being fit, and
additionally, the Nuker profile has no distinct outer edge.  To deal with these
two issues and following \cite{Tripathi2017}, we estimate the \textit{effective
radius} ($\R{eff}$), the radius at which a given fraction ($x$) of the
cumulative flux is contained.  Here, we compute two $\R{eff}$:  \R{68}, the
radius containing 68\% of the flux, to connect our results to the low- and
medium-resolution disk surveys \citep{Tripathi2017,Andrews2018a}; and \R{90},
the radius containing 90\% of the flux.  The latter is done primarily to test
how well low- and medium-resolution surveys recover disk radii obtained via
high-resolution ALMA surveys \citep{Long2018, Huang2018}, although we also find
it useful when comparing our disk sizes with dynamical features in the Solar
System (see Section \ref{sec_discussion_solar_system}).

We randomly sample 5000 models from the second half of our  MCMC chain, and for
each model we calculate \R{eff}.  This gives us a posterior distribution of
disk sizes.  We take the median value to be our \R{eff} estimate, and quantiles
of 16 and 84\% as our upper and lower confidence intervals.

In order to determine wheter the observations resolve the source or not, we
independently fit and test models created by both a Nuker profile (resolved)
and a point source (unresolved) for each observation.  The reduced chi-square
statistic of the best Nuker profile and best point source model are compared.
In cases where the point-source results in a better fit, we determine that the
source is unresolved.  To compute the upper-limit on the size of unresolved
sources, we take the results of the Nuker model, and use the 84th percentile
\R{eff} (what would be the upper confidence interval on a resolved source) as
the \R{eff} upper limit.  We expect that these disks are most likely limited by
the resolution of the telescope, but take this conservative approach in order
to not misinterpret disks that are large and faint (sensitivity limited).

Occasionally we notice walkers stuck within local minima of our posteriors.
This appears to happen in less than 15\% of our sources.  Because not all of
these stuck walkers are identifiable by eye, we make no attempt to remove them
in order to safeguard against introducing systematic errors.  However, we
measured the impact of stuck walkers on several sources and found that the decision to leave the chains unaltered
ultimately results in small changes, and slightly larger errors, in our \R{eff} estimates.
For example, an estimation of \R{68} for J16000236-4222145 with, and without
stuck walkers results in values of 
$81.67^{+1.97}_{-2.95}$
and
$85.28^{+2.62}_{-2.30}$
arcsec respectively.

\subsection{Subsample of systems with estimated disk sizes} \label{sec_modeling_sample_selection}

While we have modeled all of the detected sources within the \Lup{}, \ChaI{}, and
\USco{} regions, we do not use every source in the remainder of our analysis.

Unless otherwise noted, we have removed from our sample, disks around multiple
star systems with separations $\leq 2.0$'' in order to make our sample
consistent with those in \cite{Tripathi2017} and \cite{Andrews2018a}. These
papers imposed the separation threshold to exclude disks that might have their
sizes truncated by dynamical interactions with their companions \citep[see, e.g.][]{Manara2019}.  However, we
do not impose a flux cutoff of 2~mJy as in \cite{Andrews2018a}.

Additionally, we find 15 disks with best-fit models that we do not trust, hence
we remove them from our analysis.  The details and further discussion of the
excluded models is given in Appendix~\ref{app_sample_selection}.  This leaves
us with 152 disks: 50 from \Lup{}, 58 from \ChaI{} and 44 from \USco{} which we
include in our analysis.

To better determine if our final disk samples are representative, or biased
(being a subset of each region's complete disk population), we attempt to
reproduce previously reported disk-host (\MdustvsMstar{}) relationships with
our subset of disks (using the equivalent \LmmvsMstar\ relationship).  With the
exception of \Oph{}, this relationship has been quantified in
\cite{Ansdell2016} and \cite{Pascucci2016}, and we use the fitting slopes
reported in \cite{Pascucci2016} for our comparison given that stellar masses in
this paper are estimated in the same way.  In those works, a larger sample size
for each region (than we use here) is included in the analysis because they are
able to include flux density non-detections.  We find that for our subset of disks in each
region, \LmmvsMstar{} is correlated for \Lup{}, \ChaI{} and \USco{}, and is not
correlated for the regions \Oph{} and \Tau{} (see Appendix \ref{app_LmmvsMstar};
Figure~\ref{fig_Lmm_vs_Mstar} and Table~\ref{tab_Lmm_vs_Mstar}).  For the
regions with correlations (\Lup{}, \ChaI{} and \USco{}) we find consistently shallower
slopes as it is expected from samples lacking the faintest disks
(see Section~4 in \citealt{Pascucci2016}).
The results of our modeling for individual sources is given in Table \ref{tab_modeling}.

Appendix~\ref{app_correlation_r90_r68} shows that \R{90} and \R{68} are
strongly correlated and we derive an equation to convert  \R{68} into \R{90}
based on our modeling of the \Lup{}, \ChaI{} and \USco{} disks. Additionally,
we compare in  Appendix \ref{app_high_vs_low}  disk radii obtained here from
those obtained at high-resolution for 26 sources and demonstrate that these
lower resolution surveys are sufficient to determine dust disk sizes.

\section{Results} \label{sec_results}

We present a census of 199 homogeneously derived dust-disk sizes from five
star-forming regions and associations. In this paper,  we estimate the sizes of
152 disks: 50 from \Lup{}, 58 from \ChaI{} and 44 from \USco{}.  Of the 152
disks we model, we find 85 to be resolved, and 67 to be unresolved.  Of the unresolved disks, 20
are in \Lup{}, 25 in \ChaI{} and 22 in \USco{}\footnote{The modeling results of
individual sources are given in Appendix \ref{app_individual_modeling_results} }.  To this
we add the literature values for 20 \Oph{} disks and 27 \Tau{} disks; these
disk size estimations were originally modeled in \cite{Tripathi2017} but we use
the updated values reported in \cite{Andrews2018a}.  

Table
\ref{tab_resolved_disk_sizes} summarizes the median and maximum \R{68} and
\R{90} sizes of resolved disks within each region.  Both interpretations of \R{eff} are useful to
consider.  \R{68} is necessary in order to compare all 5 regions, since only
\R{68} sizes are available for \Oph{} and \Tau{}.  Consequently, the majority
of our analysis uses \R{68} as a proxy for \R{eff}.  However, \R{90} better
approximates the full extent of the disk,
and we find it useful to consider in the discussion (Section
\ref{sec_discussion}).


\begin{figure}[ht!]
\plotone{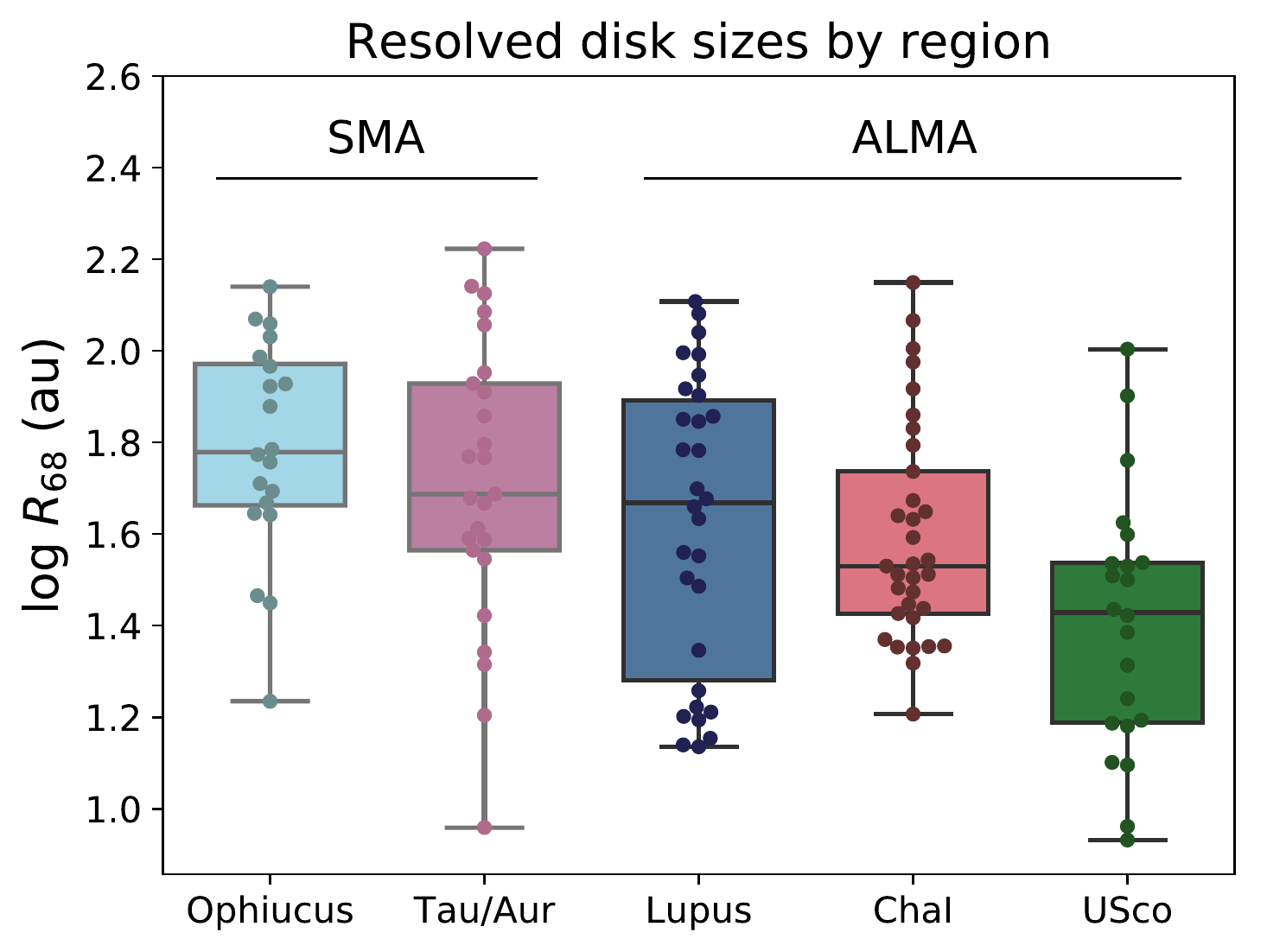}
    \caption{Swarmplots for resolved disks in different regions, ordered by age. The boxplots include a shaded region surrounding the \R{68} 25-75\% quartiles, the horizontal line denotes the median disk size, while whiskers  define the 0-25\% and 75-100\% quartiles. The regions observed with the SMA are greyed out because they are biased to the brightest millimeter disks, hence their size distributions should not be directly compared to the regions observed by ALMA.
    \label{fig_region_disk_sizes_swarm_boxplot}}
\end{figure}

A comparison of the size distribution for each region is shown in Figure
\ref{fig_region_disk_sizes_swarm_boxplot}. Individual resolved disks within
each region are shown as swarmplots with a shaded box surrounding the
distribution's 25-75\% quartiles and a horizontal line denoting the median disk
size.  Whiskers extend from the shaded boxes defining the 0-25\% and 75-100\%
quartiles. While it may be tempting to infer a trend of decreasing disk size
with age, it is important to recall that the \Oph{} and \Tau{} samples are
biased towards higher luminosity disks, and therefore, a direct comparison of
size distributions, in particular of the minimum and median size, between the
SMA and ALMA samples is unjustified. The largest disks, being also among the
brightest (see Section~5.1), are the least affected by the bias mentioned
above, as well as by differences in survey sensitivity and spatial resolution.
Table~\ref{tab_resolved_disk_sizes} and
Figure~\ref{fig_region_disk_sizes_swarm_boxplot} show that all regions, except
USco, have multiple disks with \R{68} greater than 115\,au (while USco has only one disk with \R{68} larger than 80\,au), hinting that USco disks are
smaller than those in other regions. This is in agreement with \cite{Barenfeld2017}
who used a different approach to determine USco disk sizes\footnote{\cite{Barenfeld2017}
fit power-law models to the dust surface density and carry out continuum
radiative transfer calculations to compute the surface brightness and
visibilities.} and concluded that that they are $\sim 3$ times smaller than
those in  \Oph{}, \Tau{} and the subset of the \Lup{} disks modelled by \cite{Tazzari2017}.

\begin{deluxetable}{@{\extracolsep{4pt}}lrrrrrrr}[ht]
    \tablecaption{Summary statistics for resolved disks
    \label{tab_resolved_disk_sizes}}
\tablewidth{0pt}
\tablehead{
    \colhead{} &
    \colhead{} &
    \multicolumn2c{\R{68} (au)} &
    \multicolumn2c{\R{90} (au)} \\
\cline{3-4}
\cline{5-6}
    \colhead{region} &
    \colhead{count} &
    \colhead{median} &
    \colhead{max} &
    \colhead{median} &
    \colhead{max}
}
\startdata
    \Oph{}  & 20 & 60.1 & 138.0 & ...  & ... \\
    \Tau{}  & 25 & 48.6 & 167.0 & ...  & ... \\
    \Lup{}  & 30 & 46.6 & 128.1 & 63.2 & 213.1 \\
    \ChaI{} & 33 & 33.9 & 140.9 & 43.1 & 231.3 \\
    \USco{} & 22 & 26.8 & 100.8 & 33.4 & 126.2 \\
\enddata
    \tablecomments{
        \Oph{} and \Tau{} \R{68} values are from \cite{Andrews2018a}.
    }
\end{deluxetable}

\begin{figure}[ht!]
\plotone{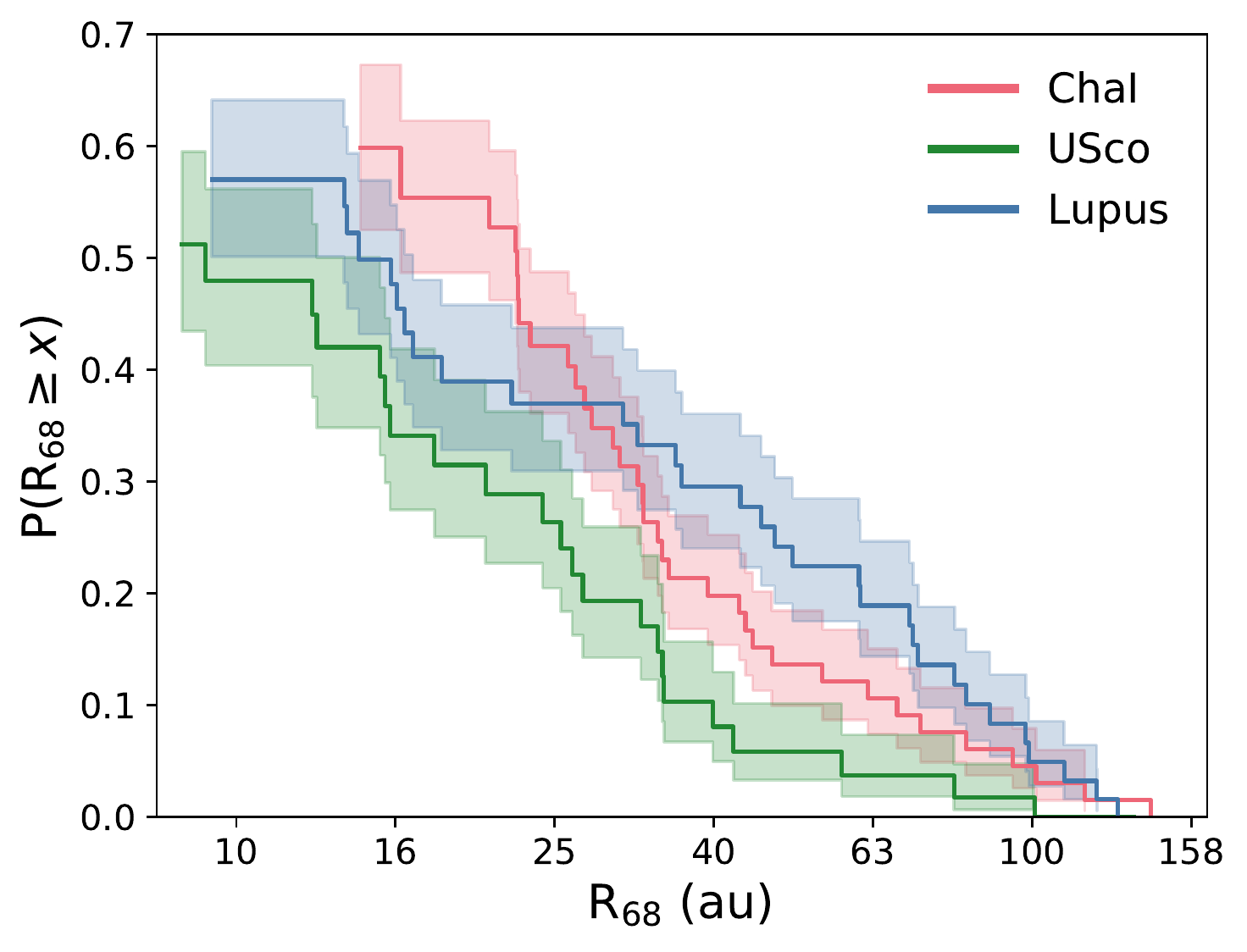}
    \caption{Cumulative disk sizes distributions for our modeled regions: \Lup{}, \ChaI{} and \USco{}.  Shaded regions indicate $1\sigma$ confidence intervals. \label{fig_survival}}
\end{figure}

To further examine if there is a difference in the observed distributions of
the now homogeneously-derived disk sizes, we focus on the three ALMA regions
observed with similar sensitivity and spatial resolution (\Lup{}, \ChaI{} and
\USco{}) and show the \R{68} cumulative distribution functions (CDF) in
Figure~\ref{fig_survival}.  Uncertainties on the CDF are determined using the
Kaplan-Meier estimator\footnote{We use the Python lifelines
\citep{cameron_davidson_pilon_2019_3240536} Kaplan-Meier implementation.} and
include unresolved disks with upper limits. However, the CDF uncertainties do
not consider our disk size uncertainties.  There are two features in
Figure~\ref{fig_survival} that are worth noting.  First, there is a deficit of
large disks in \USco{} when compared to \Lup{} and \ChaI{}.  Second, all
regions host small disks (as small as $\sim15$\,au), and \USco{} in particular
appears to have a population of even smaller disks.  However, the smaller end
of the disk-size distributions are impacted by the source distance combined
with the chosen beam size and sensitivity.  For this reason we make no
inferences regarding the smallest disks.

In order to test if two regions  are  drawn from the same empirical
distribution function, we compare each region with every other using the
Anderson-Darling test \citep{Anderson1952}.  We report the Anderson-Darling statistic and
significance level (sig.) in Table~\ref{tab_anderson_darling}.
The significance level is the level at which we cannot reject the null hypothesis that the samples are drawn from the same distribution.
The small values of $0.013$ and
$0.014$ suggest that the
\USco{} sample is unlikely to be drawn from the same parent disk-size  distribution of \Lup{} and \ChaI{}, respectively.
With a significance level of 0.1 we are not
able to determine if \Lup{} and \ChaI{} are drawn from differing distributions or not with any high
degree of confidence.  \Oph{} and \Tau{}, with their similarly biased samples
have a significance level $>25\%$, suggesting that there is no difference in
the distribution from which the two regions' brightest disks are drawn.

To test if these results depend on the selection of \R{68} or \R{90}, we
produced a \R{90} version of Figure~\ref{fig_survival}, and performed the
Anderson-Darling tests on \Lup{}, \ChaI{} and \USco{} using our \R{90} disk
sizes and found no significant change in the observed trends.

\begin{deluxetable}{@{\extracolsep{4pt}}lllr}[ht]
    \tablecaption{Comparison of \R{68} size distributions\label{tab_anderson_darling}}
\tablewidth{0pt}
\tablehead{
\multicolumn2c{\R{68} Distributions} &
\multicolumn2c{Anderson-Darling} \\
\cline{1-2}
\cline{3-4}
\colhead{Region 1} &
\colhead{Region 2} &
\colhead{stat} &
\colhead{sig.\tablenotemark{a}}
}
\startdata
\USco{} & \ChaI{} & 3.4  & $0.014$ \\
\USco{} & \Lup{}  & 3.4  & $0.013$ \\
\ChaI{} & \Lup{}  & 1.2  & $0.105$ \\
\Lup{}  & \Oph{}  & 1.2  & $0.104$ \\
\Tau{}  & \Oph{}  & -0.2 & $>0.25$ \\
\enddata
    \tablenotetext{a}{sig., also known as the error rate, indicates the significance level at which the null hypothesis that samples are drawn from the same distribution cannot be rejected.}
\end{deluxetable}

\subsection{Relations between stellar and disk properties} \label{sec_relations}

The following two subsections examine the disk size-disk luminosity (\RvsLmm{}) and
the disk size-stellar properties (\RvsMstar{} and \RvsLstar{}) relations.  In
order to determine if empirical relationships can be established, we apply
several statistical tests to the data.

We begin with the Shapiro$-$Wilk normality test \citep{Shapiro1965} to
determine if the distribution of our bivariate data is normal or not. This is
important as many correlation tests, e.g. the Pearson's r test, are based on
the assumption that the data follow a normal distribution.  The Shapiro$-$Wilk
p-value is the null hypothesis probability that the sample is normally
distributed.  In cases where the p-value is $<0.05$ we reject the null
hypothesis and conclude that the distribution is not normal.  Otherwise, we
conclude that the distribution is likely normal.  The result of this determines
which correlation test we use afterward and is described in further detail
below.

For all regions we calculate both the Pearson correlation coefficient
(hereafter Pearson r test) and Spearman's rank correlation coefficient
(hereafter Spearman $\rho$ test) and the corresponding p-values.  For both
tests, the p-value gives the probability of rejecting the null hypothesis that
there is no statistically significant relationship between the variables, more
specifically a linear relation for the Pearson's r test and a monotonic one for
the Spearman $\rho$ test.  In cases where the p-value for the chosen statistic
is $<0.05$ we consider the data to be correlated.  As mentioned above, the
Pearson's r test requires the data to be bivariate normal.   For this reason,
we use the Pearson's r test to establish if there is a linear correlation only
when our data are normally distributed, as found by the Shapiro$-$Wilk test. In
all other cases, we rely on the Spearman $\rho$ to test for the existence of a
monotonic relationship between variables.  Note that while the Spearman $\rho$
does not need the variables to be normally distributed, it does require that
they are converted into a rank-order (ordinal) data set.

A limitation of all these statistical tests is that they do not include
upper-limits and uncertainties in the assessment. As such, and for comparison
with results already reported in the literature, we might fit linear
relationships even if our
correlation tests result in a probability larger than 0.05
that the variables are not correlated.


\subsubsection{Disk radii and millimeter luminosities} \label{sec_relation_rlmm}

\begin{figure*}[ht!]
    \plotone{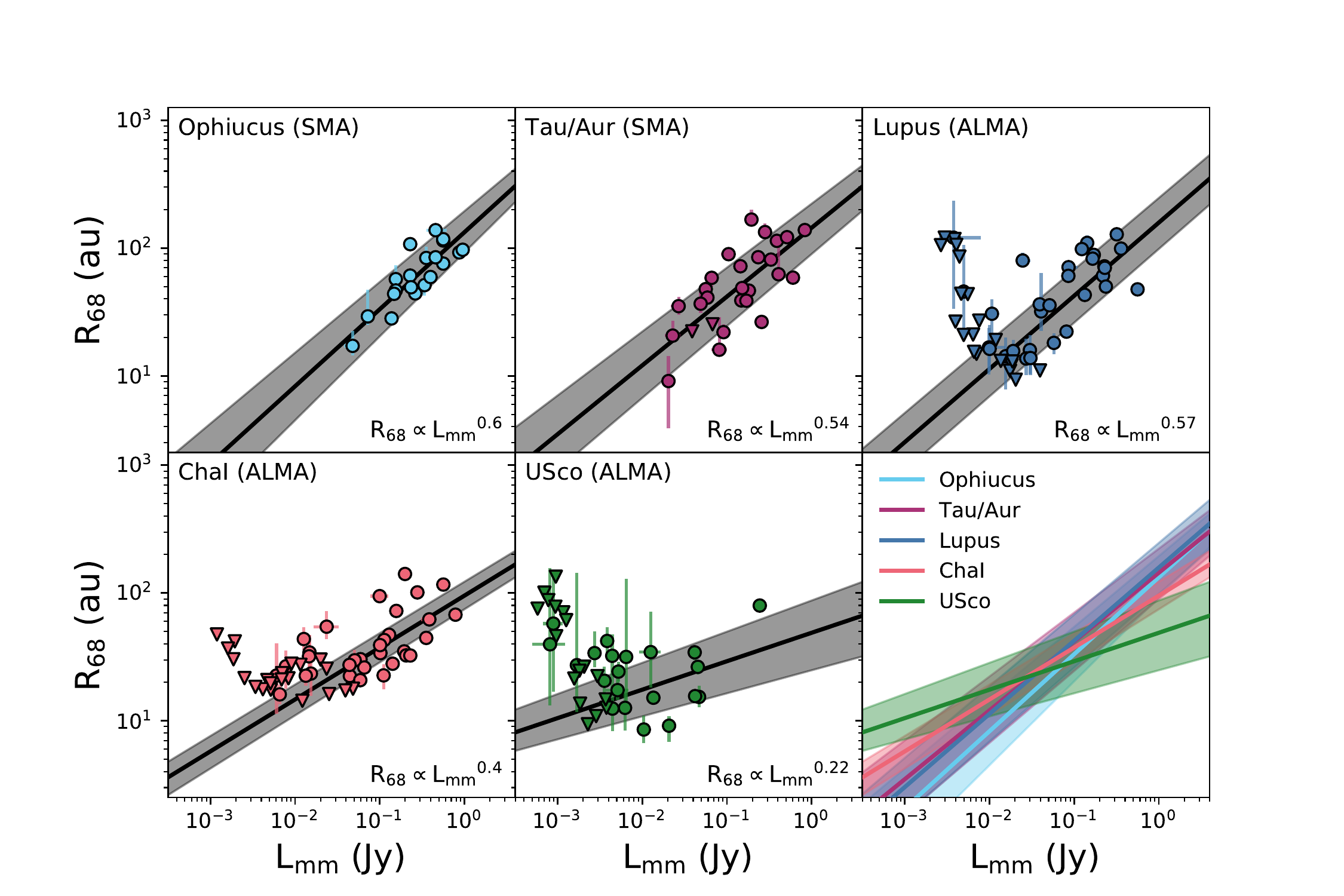}
    \caption{Fitting of \RvsLmm{}. The first 5 panels (left to right; top to bottom; ordered by region age) show the model results of each region as circles (resolved) and triangles (upper-limits).  The best fit from MCMC linear regression is plotted as a black line, and surrounded by our 68\% confidence intervals in grey.  The last panel replots the bests fits of each region (and the corresponding 68\% confidence intervals) so that they can be directly compared. Fit parameters for each region are given in Table \ref{tab_Lmm_relation}.
    \label{fig_R_vs_Lmm}}
\end{figure*}

\begin{deluxetable*}{@{\extracolsep{4pt}}lllllllllllll}[ht]
    \tablecaption{\RsixvsLmm{} Statistical tests \label{tab_Lmm_relation}}
\tablewidth{0pt}
\tablehead{
\colhead{} &
\multicolumn2c{Shapiro \Lmm{}} &
\multicolumn2c{Shapiro \R{eff}} &
\multicolumn2c{Pearson r} &
\multicolumn2c{Spearman $\rho$} & 
\multicolumn4c{Regression Parameters} \\
\cline{2-3}
\cline{4-5}
\cline{6-7}
\cline{8-9}
\cline{10-13}
\colhead{Region} &
\colhead{stat} &
\colhead{p-value} &
\colhead{stat} &
\colhead{p-value} &
\colhead{stat} &
\colhead{p-value} &
\colhead{stat} &
\colhead{p-value} &
\colhead{$\alpha$}  & 
\colhead{$\beta$}  & 
\colhead{$\sigma$} &
\colhead{$\hat{\rho}$}
}
\startdata
    \Oph{}{}  & 0.91 & 5.7e-02 & 0.96 & 5.9e-01 & 0.67       & 1.3e-03       & 0.75  & 1.4e-04 & $2.11^{+0.06}_{-0.06}$ & $0.60^{+0.11}_{-0.11}$ & $0.14^{+0.03}_{-0.02}$ & $0.92^{+0.04}_{-0.06}$\\
    \Tau{}{}  & 0.84 & 9.9e-04 & 0.91 & 3.4e-02 & \dem{0.59} & \dem{2.0e-03} & 0.70  & 1.0e-04 & $2.16^{+0.11}_{-0.11}$ & $0.53^{+0.12}_{-0.12}$ & $0.24^{+0.05}_{-0.04}$ & $0.86^{+0.05}_{-0.08}$\\
    \Lup{}{}  & 0.81 & 8.8e-05 & 0.91 & 1.9e-02 & \dem{0.44} & \dem{1.4e-02} & 0.55  & 1.8e-03 & $2.20^{+0.14}_{-0.13}$ & $0.57^{+0.10}_{-0.10}$ & $0.30^{+0.06}_{-0.05}$ & $0.88^{+0.04}_{-0.05}$\\
    \ChaI{}{} & 0.75 & 4.5e-06 & 0.77 & 1.3e-05 & \dem{0.53} & \dem{1.7e-03} & 0.64  & 7.7e-05 & $1.97^{+0.08}_{-0.08}$ & $0.40^{+0.06}_{-0.06}$ & $0.21^{+0.03}_{-0.03}$ & $0.90^{+0.03}_{-0.04}$\\
    \USco{}{} & 0.45 & 7.7e-08 & 0.87 & 1.0e-02 & \dem{0.60} & \dem{4.1e-03} & -0.27 & 2.4e-01 & $1.68^{+0.23}_{-0.22}$ & $0.22^{+0.11}_{-0.10}$ & $0.28^{+0.07}_{-0.05}$ & $0.67^{+0.12}_{-0.18}$\\
\enddata
    \tablecomments{Values we consider unreliable are greyed out (see Section~\ref{sec_relations}).
    }
\end{deluxetable*}

Recently, \cite{Andrews2018a} demonstrated that the dust disk effective radius
(\R{eff}) and the millimeter luminosity (\Lmm ) scale in the same way for their
\Oph{}, \Tau{}, and \Lup{} samples as \RpropL{0.5}.  Thus, one might infer that
such a relationship is universal and apply to all star-forming regions.  Here,
we demonstrate that this is not the case and that the \RvsLmm{} slope flattens
moving from the younger to the older star-forming regions. 

Figure \ref{fig_R_vs_Lmm} shows the inferred disk sizes (circles) or upper
limits (downward triangles) as a function of \Lmm . Values for \Oph{} and
\Tau{} are from \cite{Andrews2018a} while for the other three 
regions are from this work (see Sections \ref{sec_sample_lit} and
\ref{sec_sample_alma}). For all regions, we see the general trend of larger
disk sizes for brighter disks, although USco covers a smaller range in \Lmm{}
than other regions and the scatter is large. Indeed, when we apply the
non-parametric Spearman's rank correlation test to the disks with a measured
dust radius (circles), we find positive values (\R{eff} increases with \Lmm )
and probabilities (p-value) lower than 5\% that the two quantities are
uncorrelated in all regions except USco (see Table~\ref{tab_Lmm_relation}). 

Next, following \cite{Andrews2018a}, we fit the \RvsLmm{} relation in each
individual region taking into account measured \R{eff}, the associated
uncertainties, as well as upper limits in \R{eff}. For this task we use the
Bayesian method of linear regression described in \cite{Kelly2007} (as
implemented in the \texttt{linmix} code by Joshua E.  Meyers) and specifically
fit the following linear relation:

\begin{equation} \label{eq_reff_lmm}
    \log{\R{eff}} = \alpha + \beta \log{\Lmm{}}
\end{equation}

where $\alpha$ and $\beta$ are the intercept and the slope, respectively.  The
best fit parameters for each region, together with the scatter of the
relation ($\sigma$) and the correlation coefficient ($\hat{\rho}$), are
reported in Table~\ref{tab_Lmm_relation}. 

We find that $\hat{\rho}$, which is estimated accounting for uncertainties and
upper limits (but assumes bivariate-normal data), is positive and larger than
the Spearman $\rho$ correlation coefficient in all regions.  This may be
particularly important for \USco{}, where the Spearman $\rho$ coefficient of
-0.27 and large p-value suggest no \RvsLmm{} correlation for sources with
measured disk sizes, while the \texttt{linmix} $\hat{\rho}$ value of 0.67
points to a positive correlation, albeit less strong than in the other younger
star-forming regions.  The first five panels of Figure \ref{fig_R_vs_Lmm}
visualize the best fit for each region with the grey shadowing highlighting the
68\% confidence intervals. The sixth panel of Figure \ref{fig_R_vs_Lmm}
summarizes the results and emphasizes the main finding of our analysis that the
\RvsLmm{} relation is not universal.

We do not perform a fitting for the combined sample of all regions due to
differences in the estimated disk size uncertainties between the SMA and the
ALMA samples and our finding that not all regions share the same \RvsLmm{}
relation.


\subsubsection{Disk radii and stellar properties} \label{sec_relation_rmstar}

Several works have pointed out that \Lmm{}, which (if optically thin) probes the dust disk mass, is
positively correlated with stellar properties like stellar bolometric
luminosity (\Lstar) and stellar mass (\Mstar) in most star-forming regions,
\cite[e.g.,][]{Andrews2013,Ansdell2016,Pascucci2016}.  However, the
relationship between \R{eff} and these same stellar properties has been less
explored.  \cite{Andrews2018a} find \RvsMstar{} and \RvsLstar{} to be correlated
($\R{eff} \propto \Mstar{}^{0.6} \propto \Lstar{}^{0.3}$) for the combined
Tau/Aur and Lupus samples but to a lesser degree than \RvsLmm{}.  In this
section we investigate whether such relations are present in individual
regions.

\begin{figure*}[ht!]
    \plotone{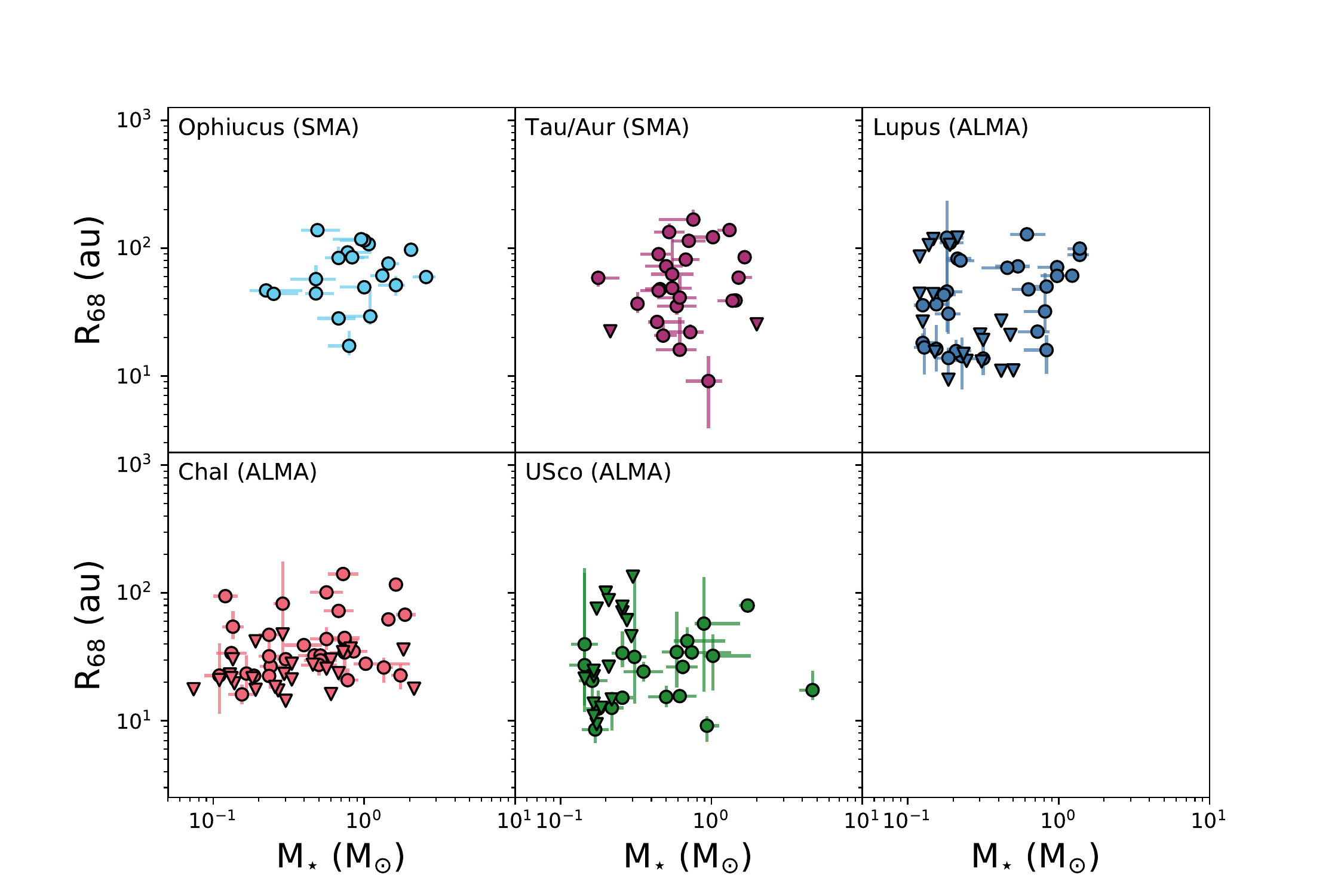}
    \caption{Fitting of \RvsMstar{}. The first 5 panels (left to right; top to bottom; ordered by region age) show the model results of each region as circles (resolved) and triangles (upper-limits).\label{fig_R_vs_Mstar}}
\end{figure*}

\begin{figure*}[ht!]
    \plotone{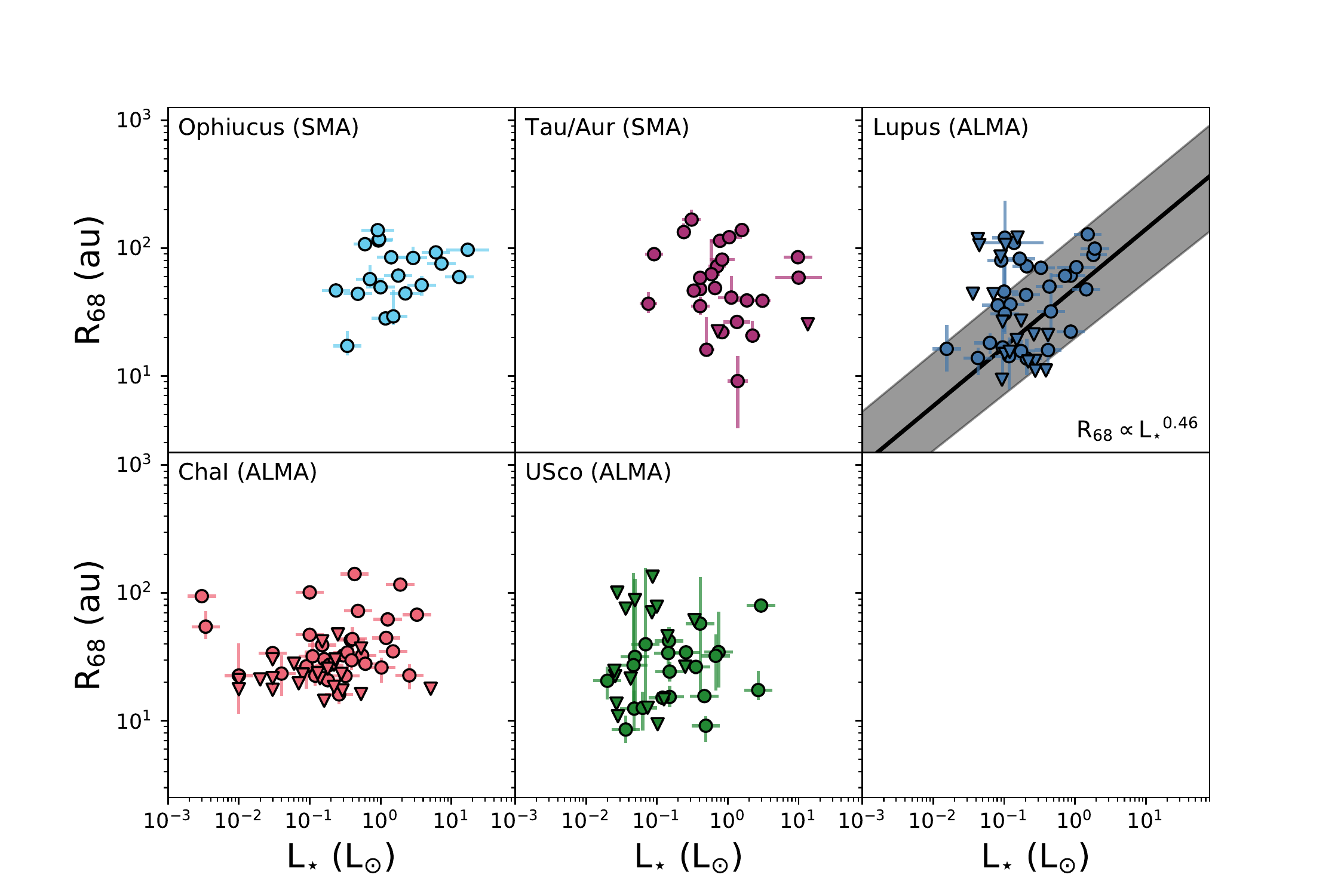}
    \caption{Fitting of \RvsLstar{}. The first 5 panels (left to right; top to bottom; ordered by region age) show the model results of each region as circles (resolved) and triangles (upper-limits).\label{fig_R_vs_Lstar}}
\end{figure*}

Figures~\ref{fig_R_vs_Mstar} and ~\ref{fig_R_vs_Lstar} show the inferred disk
sizes (circles) or upper limits (downward triangles) as a function of \Mstar{}
and \Lstar{}, respectively.  Tables~\ref{tab_Mstar_relation}  and
\ref{tab_Lstar_relation} report the statistical tests described in Section
\ref{sec_relations} as well as the \texttt{linmix} best fit parameters and
correlation coefficient $\hat{\rho}$.  For the two regions that are most biased
to bright millimeter disks (\Oph{} and \Tau{}), we can confidently conclude
that there is no correlation in \RvsMstar{} nor in \RvsLstar{}.

The three remaining regions (\Lup{}, \ChaI{}, and \USco{}) have Spearman $\rho$
correlation coefficients (0.29-0.34) and p-values (0.07-0.20) where we can
not completely rule out a weak or marginal correlation between $R_{eff}$ and
$\Mstar$, but correlations are not strongly supported.

Interestingly, for \ChaI{}, which covers the largest range in \Lstar{}, we can
rule out a \RvsLstar{} correlation with a high level of confidence (correlation
coefficient=0.17 and p-value=0.35).

The situation is different for \USco{} where we can not rule out a weak
\RvsLstar{} relation, the Spearman $\rho$ correlation coefficient is 0.29 with
a p-value of 0.19 and $\hat{\rho}$ is 0.72.

Finally, the \Lup{} region appears to have some degree of correlation in
\RvsLstar{}, but likely not in \RvsMstar{}.  When we combine the \Oph{},
\Tau{}, and \Lup{} samples and refit the \RvsMstar{} relation with
\texttt{linmix} we find $\hat{\rho}=0.63^{+0.07}_{-0.09}$, the same as for
Lupus. Thus, we conclude that the weak \RvsMstar{} correlation reported by
\cite{Andrews2018a} for the SMA+ALMA samples (mostly \Oph{}, \Tau{}, and
\Lup{}) with a coefficient of $\hat{\rho}=0.54$ is mostly driven by the Lupus
sample.   Similarly, when we combine \Oph{}, \Tau{}, and \Lup{} samples and
refit the \RvsLstar{} relation, we find a coefficient of
$\hat{\rho}=0.64^{+0.07}_{-0.09}$, consistent with the findings in
\cite{Andrews2018a} for the same joint sample.

\begin{deluxetable*}{@{\extracolsep{4pt}}lllllllllllll}[ht]
    \tablecaption{\RsixvsMstar{} statistical tests \label{tab_Mstar_relation}}
\tablewidth{0pt}
\tablehead{
\colhead{} &
\multicolumn2c{Shapiro \Mstar{}} &
\multicolumn2c{Shapiro \R{eff}} &
\multicolumn2c{Pearson r} &
\multicolumn2c{Spearman $\rho$} & 
\multicolumn4c{Regression Parameters} \\
\cline{2-3}
\cline{4-5}
\cline{6-7}
\cline{8-9}
\cline{10-13}
\colhead{Region} &
\colhead{stat} &
\colhead{p-value} &
\colhead{stat} &
\colhead{p-value} &
\colhead{stat} &
\colhead{p-value} &
\colhead{stat} &
\colhead{p-value} &
\colhead{$\alpha$}  & 
\colhead{$\beta$}  & 
\colhead{$\sigma$} &
\colhead{$\hat{\rho}$}
}
\startdata
    \Oph{}  & 0.91 & 6.7e-02 & 0.96 & 5.9e-01 & \dem{0.08} & \dem{7.2e-01} & 0.24 & 3.2e-01 & \dem{$1.81^{+0.06}_{-0.06}$} & \dem{$0.20^{+0.25}_{-0.24}$} & \dem{$0.24^{+0.05}_{-0.04}$} & $0.47^{+0.22}_{-0.11}$\\
    \Tau{}  & 0.87 & 3.7e-03 & 0.91 & 3.4e-02 & \dem{0.16} & \dem{4.3e-01} & 0.18 & 3.8e-01 & \dem{$1.71^{+0.08}_{-0.09}$} & \dem{$0.18^{+0.30}_{-0.29}$} & \dem{$0.34^{+0.07}_{-0.05}$} & $0.38^{+0.23}_{-0.10}$\\
    \Lup{}  & 0.82 & 2.3e-04 & 0.91 & 1.6e-02 & \dem{0.31} & \dem{1.0e-01} & 0.34 & 7.2e-02 & \dem{$1.66^{+0.13}_{-0.13}$} & \dem{$0.57^{+0.23}_{-0.22}$} & \dem{$0.43^{+0.08}_{-0.06}$} & $0.62^{+0.10}_{-0.14}$\\
    \ChaI{} & 0.85 & 4.0e-04 & 0.80 & 2.7e-05 & \dem{0.23} & \dem{2.0e-01} & 0.23 & 2.0e-01 & \dem{$1.50^{+0.07}_{-0.08}$} & \dem{$0.31^{+0.14}_{-0.14}$} & \dem{$0.33^{+0.06}_{-0.04}$} & $0.58^{+0.11}_{-0.15}$\\
    \USco{} & 0.56 & 7.4e-07 & 0.87 & 1.0e-02 & \dem{0.13} & \dem{5.8e-01} & 0.29 & 1.9e-01 & \dem{$1.41^{+0.08}_{-0.08}$} & \dem{$0.47^{+0.16}_{-0.15}$} & \dem{$0.26^{+0.06}_{-0.05}$} & $0.72^{+0.09}_{-0.13}$\\
\enddata
    \tablecomments{Values we consider unreliable are greyed out (see Section~\ref{sec_relations}).}
\end{deluxetable*}


\begin{deluxetable*}{@{\extracolsep{4pt}}lllllllllllll}[ht]
    \tablecaption{\RsixvsLstar{} statistical tests \label{tab_Lstar_relation}}
\tablewidth{0pt}
\tablehead{
\colhead{} &
\multicolumn2c{Shapiro \Lstar{}} &
\multicolumn2c{Shapiro \R{eff}} &
\multicolumn2c{Pearson r} &
\multicolumn2c{Spearman $\rho$} & 
\multicolumn4c{Regression Parameters} \\
\cline{2-3}
\cline{4-5}
\cline{6-7}
\cline{8-9}
\cline{10-13}
\colhead{Region} &
\colhead{stat} &
\colhead{p-value} &
\colhead{stat} &
\colhead{p-value} &
\colhead{stat} &
\colhead{p-value} &
\colhead{stat} &
\colhead{p-value} &
\colhead{$\alpha$}  & 
\colhead{$\beta$}  & 
\colhead{$\sigma$} &
\colhead{$\hat{\rho}$}
}
\startdata
    \Oph{}  & 0.66 & 1.3e-05 & 0.96 & 5.9e-01 & \dem{0.15}  & \dem{5.4e-01} & 0.17  & 4.6e-01 & \dem{$1.77^{+0.06}_{-0.06}$} & \dem{$0.14^{+0.13}_{-0.13}$}  & \dem{$0.24^{+0.05}_{-0.04}$} & $0.54^{+0.19}_{-0.38}$\\
    \Tau{}  & 0.54 & 9.5e-08 & 0.91 & 3.4e-02 & \dem{-0.02} & \dem{9.3e-01} & -0.18 & 3.9e-01 & \dem{$1.67^{+0.07}_{-0.07}$} & \dem{$-0.15^{+0.13}_{-0.14}$} & \dem{$0.33^{+0.06}_{-0.05}$} & \dem{$nan^{+nan}_{-nan}$}\\
    \Lup{}  & 0.75 & 1.4e-05 & 0.91 & 1.6e-02 & \dem{0.45}  & \dem{1.4e-02} & 0.42  & 2.3e-02 & $1.68^{+0.13}_{-0.13}$ & $0.46^{+0.17}_{-0.16}$  & $0.41^{+0.08}_{-0.06}$ & $0.67^{+0.10}_{-0.13}$\\
    \ChaI{} & 0.71 & 1.2e-06 & 0.77 & 1.3e-05 & \dem{0.18}  & \dem{3.2e-01} & 0.17  & 3.5e-01 & \dem{$1.45^{+0.08}_{-0.09}$} & \dem{$0.10^{+0.09}_{-0.08}$}  & \dem{$0.36^{+0.06}_{-0.05}$} & $0.45^{+0.15}_{-0.24}$\\
    \USco{} & 0.57 & 8.9e-07 & 0.87 & 1.0e-02 & \dem{0.45}  & \dem{3.9e-02} & 0.35  & 1.2e-01 & \dem{$1.49^{+0.10}_{-0.10}$} & \dem{$0.33^{+0.11}_{-0.10}$}  & \dem{$0.25^{+0.06}_{-0.05}$} & $0.77^{+0.08}_{-0.12}$\\
\enddata
    \tablecomments{Values we consider unreliable are greyed out (see Section~\ref{sec_relations}), including the result of $nan$ which denotes a value below the floating point precision of our analysis.}
\end{deluxetable*}

\section{Discussion} \label{sec_discussion}

    In the following sections we compare
    the size distributions found in each region (Section
    \ref{sec_discussion_regions}), define and compare our inferred disk sizes with the
    outer edge of the Solar System (Section \ref{sec_discussion_solar_system}),
    and discuss disk-disk and disk-host scaling relations (Section
    \ref{sec_discussion_scaling_relations}).

    \subsection{Comparison between regions} \label{sec_discussion_regions}

        In Section \ref{sec_results} we use several statistical approaches
        to assess how similar, or different, the distributions of disk sizes are in our
        sample regions.

        When compared to other regions \USco{} lacks large dust disks.  The vast
        majority ($\ge 75$\%) of resolved disks within \USco{} have sizes that fall
        below the \Lup{} and \ChaI{} median disk sizes.  \cite{Barenfeld2017} compared
        their \USco{} disk sizes with the inhomogeneously derived sizes of \Oph{},
        \Tau{} and \Lup{} \footnote{The comparison by \cite{Barenfeld2017} made use of
        \Lup{} dust-disk sizes from \cite{Tazzari2018} which considered a sample that
        excluded edge-on disks, disks with sub-structures and disks with mm-flux $<
        4\text{mJy}$, resulting in $\sim$\,50\% of the objects modelled in
        \cite{Andrews2018a} or this work.} and concluded that \USco{} disks are three
        times smaller.  We also find that the typical disk in \USco{} is smaller than
        the typical disk  in \Lup{} or \ChaI{}.  However, the difference is not as
        great: the median disk size in \USco{} is 1.7 times smaller than that in \Lup{}
        and 1.3 times smaller than that in \ChaI{}.  We believe that the difference
        between our results and \cite{Barenfeld2017} is due to the fact that the latter
        has used inhomogeneous disk size estimations as well as samples biased to the
        brightest disks in \Oph{}, \Tau{}, and \Lup{}.  Because the disks from \Oph{}
        and \Tau{} are biased towards only the brightest disks, the only conclusion we
        can make is that \USco{} lacks the large disks seen in both of those regions.
        The decrease in median disk size from the younger \Lup{} and \ChaI{} to the
        older \USco{} may be interpreted as an evolution of the disk outer edge caused by e.g. efficient inward drift of millimeter grains \citep{Pinilla2013, Krijt2015, Pascucci2016}; growth of millimeter grains into planetesimals \citep{Barenfeld2017, Gerbig2019, Lenz2019}; or external photoevaporation in the higher UV field of the USco OB association \citep{Facchini2016, Barenfeld2017}.
        Measuring gas disk sizes in these three regions should help to discriminate between internal vs external processes. 

        \cite{Eisner2018} reported that the population of disk sizes within \ChaI{} was
        significantly smaller than that of disks in the younger regions of \Oph{},
        \Tau{}, and \Lup{}.  However, we do not arrive at the same conclusion.  As
        previously mentioned, we find that \ChaI{} disks are not significantly smaller
        than \Lup{}, and span a similar range in disk sizes.  \ChaI{}, when compared to
        \Lup{} in Figure \ref{fig_survival}, hints at a enhanced population of disks
        between 15 and 25\,au, and a decreased population of disks between 30 and
        65\,au.

        The differing result from \cite{Eisner2018} is likely due to
        several factors.  First, their work compares inhomogeneously estimated
        disk sizes, e.g. FWHMs from \cite{Pascucci2016} for Cha~I with
        exponential cutoff radii of a power-law disk from Tazzari et al. (2017)
        for Lupus.  Second, the entire \ChaI{} disk population is compared with
        luminosity biased samples in \Oph{}, \Tau{}, and even Lupus (only the
        sub-sample in \cite{Tazzari2017} was available at that time).   Lastly, the cumulative disk-size
        distributions used for comparison in \cite{Eisner2018} are not
        consistently constructed, leading to a different definition for what a
        probability of unity means in each region. For instance,  the \ChaI{}
        sample appears to include all sources that were targeted in
        \cite{Pascucci2016}, whether they were detected or not,  the \USco{}
        sample includes only resolved sources, while the \Lup{} sample appears
        to include detected sources, whether they were resolved or not.
        
    \subsection{The outer edge of the Solar System} \label{sec_discussion_solar_system}

    Stellar encounters \citep[e.g.][]{Ida2000,Kenyon2004} and external photoevaporation
    by massive stars within a star cluster \citep[e.g.][]{Adams2004} have been
    suggested as mechanisms connected to the formation of the Solar System's
    outer edge.  In this section we look into whether or not our results can
    test these models which utilize edge truncation  to explain the
    size of the Solar System.

    The region of the Solar System beyond Neptune (the trans-Neptune region or
    Edgeworth--Kuiper belt) is populated with icy bodies (TNOs) which fall into
    distinct classes based on their dynamical properties. 
    There is a decrease in the population of TNOs beyond 48\,au, and the
    population of objects with nearly circular orbits effectively ends at 45~au
    \citep[e.g.][]{Petit2011}.  This \textit{outer edge} of the Kuiper belt
    appears well defined and is not an observational bias
    \citep{Allen2001,Morbidelli2008}.
    Because the region beyond this edge
    is dynamically stable on timescales longer than the Solar System's lifetime
    \citep{Duncan1995}, a primordial population of planetesimals beyond 45\,au
    -- had it existed -- should have been retained.

    Determining if the outer edge of the Solar System's primordial planetesimal
    disk was located at, beyond, or interior to the present Kuiper belt's edge
    would test hypotheses which evoke an outer-edge modifying event.  This
    work does not directly probe the history of the Solar System. However, if
    we take the location of dust observed in our disks to be an indicator of
    the location of planetesimals (or planetesimal formation), and if we
    interpret the outer edge at 45-48\,au to be primordial, we can compare the
    Solar System's outer edge with our inferred disk sizes.  It is
    important to be aware that this comparison is between the location of dust
    emission in 1-3\,Myr old disks and the location of a dynamical and
    occurrence-rate feature found within a distribution of 4.5\,Gyr old
    planetesimals.  The validity of this comparison ultimately relies on
    whether or not the Solar System's outer-edge is a feature inherited from
    its protoplanetary disk.
    
    \begin{figure}[t]
        \plotone{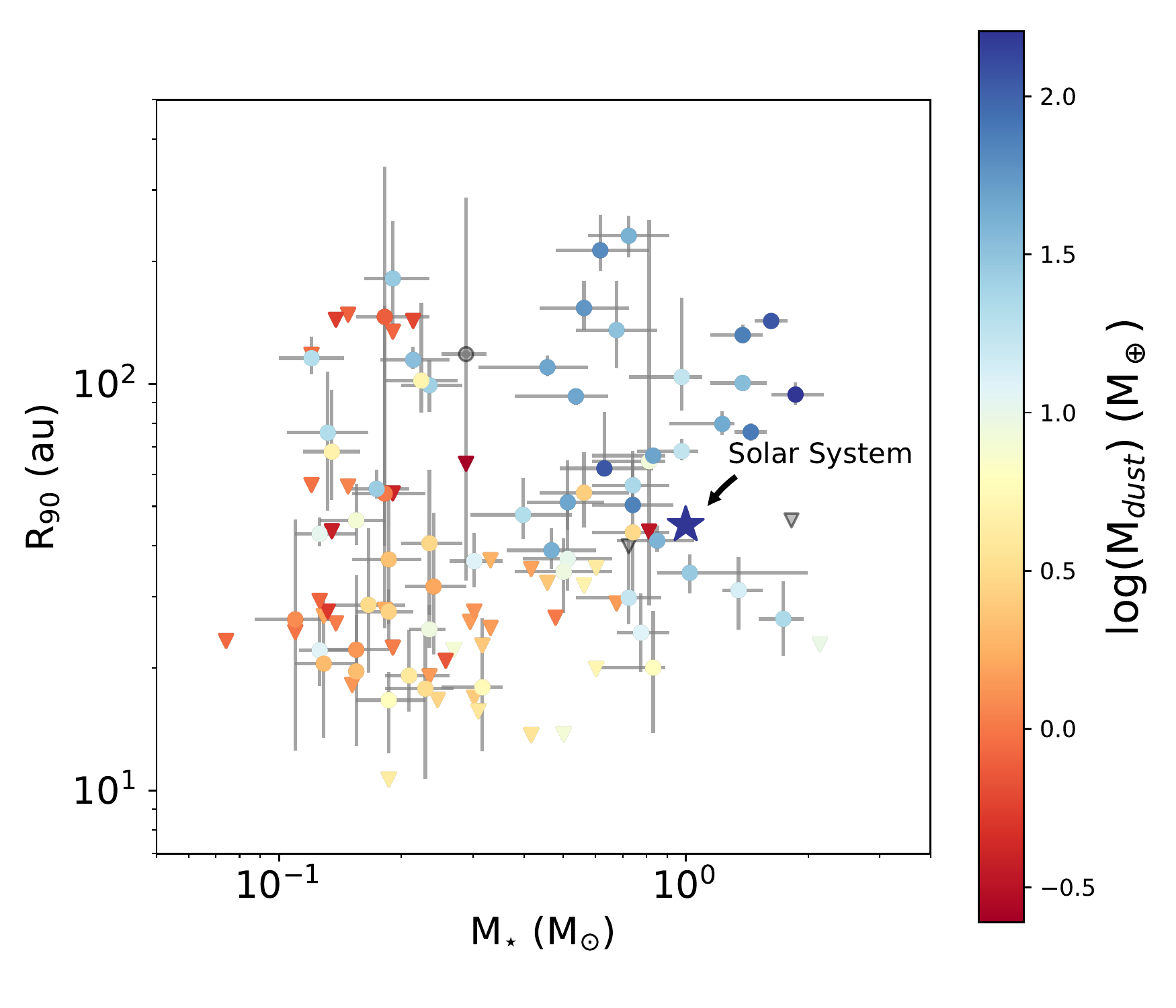}

        \caption{
            The \R{90} sizes and dust masses for disks in the 1-3 and 2-3\,Myr old
            \Lup{} and \ChaI{} star-forming regions are plotted along with the
            Solar System in order to determine if the Solar System (denoted in
            the figure with a star) should be considered \textit{typical}, or a
            statistical outlier.  Disks with constrained size estimates are
            shown as circles, and upper limits with triangles.  The color of
            each symbol (including the Solar System) corresponds to the dust
            mass.  \label{fig_solar_system}}

    \end{figure}

    While stellar encounters and external photoevaporation by massive stars may
    impact disk sizes in high-mass star-forming regions, these two mechanisms
    are most likely not affecting the disk sizes in \Lup{} and \ChaI{}.
    \cite{Winter2018} compared the effect of external photoevaporation and
    close stellar encounters on disk sizes in typical cluster
    environments and concluded that
    tidal truncation due to stellar encounters are unlikely.
    They
    find that significant truncation due to stellar encounters requires a
    cluster stellar density of $5 \times 10^4$pc$^{-3}$; far larger than the
    current stellar densities of \Lup{}
    \citep{Nakajima2000,Merin2008,Winter2018} ($<500$pc$^{-3}$) and \ChaI{} \citep{Sacco2017}.
    By comparing Gaia observations of the structural properties of \ChaI{} with $N$-body simulations,
    \cite{Sacco2017} concluded that \ChaI{} likely did not form in a
    high-density environment.  The number of O- and B-
    stars in \Lup{} and \ChaI{} is only 2 \citep{Comeron2008} and 3
    \citep{Luhman2008}, respectively, ruling out external photoevaporation as
    the dominant mechanism setting disk sizes in these low-mass star-forming
    regions.

    The Solar System, on the other hand, may have formed in a high-density
    cluster environment.
    The review by \cite{Adams2010} argues for a large cluster ($N=10^3-10^4$) as birth
    environment, and required a nearby supernova 
    to produce the inventory of short-lived radio isotopes found
    in the meteoritic record at the time of writing.  However,
    more recently \cite{Wasserburg2017} showed that the short-lived
    radionuclide abundances of $^{26}$Al, $^{60}$Fe, $^{182}$Hf, and $^{107}$Pd
    found in meteorites are not consistent with being injected via sources of
    mass $>5$\,M$_{\odot}$ such as supernovae.  A small fraction of
    presolar grains (X grains) may still link the early Solar System
    formation to a nearby supernova event \citep[see][]{Zinner2014}, but how
    nearby the supernova would need to be is not known.  Low-mass star-forming
    regions such as \Lup{} and \Oph{} are known to have been influenced by
    external Supernovae \citep[e.g.][]{Comeron2008,Wilking2005}. Therefore, it
    remains unclear if the presence of X grains limits the Solar System birth
    environment to that of a high-density cluster.

    To compare our dust-disk sizes with the Solar System, we utilize the \R{90}
    estimates which better represent the full extent of the disk. In addition,
    we constrain ourselves to the younger and lower mass star-forming regions
    of  \Lup{} and \ChaI{} to exclude significant evolution and external
    processes affecting disk sizes.  Figure \ref{fig_solar_system} provides
    this comparison and shows that the Solar System's size falls within the
    range of \R{90} values found for both low-mass star-forming regions.  We
    also test if the dust masses of the \Lup{} and \ChaI{} disks is consistent,
    or discordant, with the Solar System.  To get a rough estimate of the mass
    of the primordial Solar System's dust disk, we sum the masses of solids
    locked up within the major planets ($\sim30\,M_\oplus$).  Dust disk masses
    are estimated assuming optically thin emission and a fixed temperature of
    20\,K for the emitting grains as in the 20\,K dust-mass calculation
    described in \cite{Pascucci2016}.  Figure \ref{fig_solar_system} shows that
    the mass of solids in the Solar System is larger than the average dust disk
    mass but still falls within the range of masses observed in other disks.
    
    When compared to the disks in \Lup{} and \ChaI{}, the Solar System doesn't
    appear to be an outlier.  It is neither small, nor does it appear to be
    missing dust mass.  If the initial sizes of protoplanetary disks within
    high-density and low-density regions are similar, our results show that
    the Solar system's primordial disk requires no external modification (e.g.
    stellar encounter or photoevaporation) to explain its size and mass since it is
    consistent with a population of disks lacking truncation. 
    Radial drift of dust grains, which is common to all environments, may be setting disk sizes. 
    However, if the Solar System formed in a high-density cluster, \textit{and}
    the initial sizes of these disks is not similar to the initial sizes found
    in low-density regions, we can come to no conclusion about the history of
    the Solar System's outer edge.

    \subsection{Disk size scaling relations} \label{sec_discussion_scaling_relations}

    In Section \ref{sec_relations} we test for the existence of and, if found,
    quantify empirical relations between disk sizes ($R_{\rm eff}$), millimeter luminosities (\Lmm ), stellar masses (\Mstar ), 
    and stellar luminosities (\Lstar ).  In this sub-section we discuss our findings in that order,
    starting with \RvsLmm{}.


    We find that, in log scale, disk radii are linearly correlated with
    millimeter luminosities and that the slope of the \RvsLmm{} relation is
    not the same for all regions investigated here (see Figure
    \ref{fig_R_vs_Lmm} and Table \ref{tab_Lmm_relation}).  The latter result is different from earlier works.  Previous analysis by
    \cite{Tripathi2017} and \cite{Andrews2018a} found a slope of $~\sim\,0.5$
    to be common to the \Oph{}, \Tau , and \Lup{}  star-forming regions.  Using
    a different technique, \cite{Barenfeld2017} showed that disk sizes
    in USco, while being a factor of $\sim$3 times smaller than those in \Oph{}
    and \Tau , plot on the same \RvsLmm{} relation with slope 0.5. This
    result has led to speculation that the \RvsLmm{} relationship could be
    universal \citep[e.g.,][]{Rosotti2019}. As the relation implies that the
    millimeter luminosity scales with the square of the dust radius, the most
    commonly adopted interpretation is that the millimeter emission is mostly
    optically thick.
    
    Assuming a simple parameterization for the dust temperature profile,
    \cite{Andrews2018a} considered whether their \RvsLmm{} and \LstarvsLmm{}
    relationships for \Oph{}, \Tau{} and \Lup{} were consistent with disks
    being optically thin or thick. Ultimately, they concluded that neither of
    the two scenarios could be ruled out. Here, we focus on the
    optically thick scenario and repeat \cite{Andrews2018a} approach for
    \ChaI{} and \USco . Using the following temperature profile:  

    \begin{equation}\label{eq_temperature}
    T_d(r) \propto \left(\frac{\Lstar}{\Lsun}\right)^{0.25}\left(\frac{r}{r_0}\right)^{-q}
    \end{equation}
    
    where $r_0=10$\,au for a solar luminosity star,
    \cite{Andrews2018a} derived a relationship connecting the dust effective radius to the stellar luminosity and millimeter luminosity via the power law index ($q$)
    of the  dust temperature profile:   
    \begin{equation}\label{eq_reff_star}
        \R{eff} \propto \Lstar{}^{-1/3(2-q)}\Lmm{}^{1/(2-q)}
    \end{equation}

    We can then estimate $q$ for each region using Equation \ref{eq_reff_star}
    and our empirically measured \LstarvsLmm{} and \RvsLmm{} relationships (see
    Table \ref{tab_Lmm_vs_Lstar} in Appendix \ref{app_LmmvsLstar} and Table
    \ref{tab_Lmm_relation} in Section \ref{sec_relation_rlmm} respectively).
    With an estimate of $q$ in hand, we can test if the resulting temperature
    profile is consistent with that expected for optically thin or thick disks.
    Doing so results in values of $q$ of $0.32^{+0.33}_{-0.4}$ for \ChaI{} and
    $-0.85^{+1.42}_{-3.1}$ for \USco{}. At long millimeter wavelengths and in
    the case of optically thin emission, the radial temperature profile can be
    expressed as $T(r) \propto r^{-2/(4+\beta)}$ where $\beta$ is the index of
    the dust absorption coefficient, e.g. \cite{Evans1994}. Thus, for the average $\beta$
    of 0.6 \citep{Ricci2010}, we find $T(r) \propto r^{-0.4}$ in the optically
    thin regime.  Partially optically thick emission is characterized by a
    steeper dependence with radial distance \citep[e.g., Fig. 5 in
    ][]{Pascucci2004} close to $r^{-0.5}$ for the midplane of an accretion disk
    irradiated by the central star \citep{Dalessio1998}.  The upper bounds of
    the confidence intervals encompass  the optically thin and thick
    temperature profiles  both for \ChaI{} and \USco{}.  However, the mostly
    negative values of $q$ for \USco{} imply  a temperature profile increasing
    with disk radius, which is highly unlikely, suggesting that the disk
    emission is not optically thick.

    \cite{Andrews2018a} report finding weak scaling relationships for their joint
    sample of \Oph{}, \Tau{} and \Lup{} for both \RvsMstar{} and \RvsLstar{}.  We
    determine that \RvsMstar{} is not correlated for any of our regions, and for
    the \RvsLstar{} correlation we find a trend only for  \Lup{}. Extending disk
    radii measurements to even smaller and less luminous stars could reveal trends
    also in other regions. This would require observations more sensitive than
    those adopted for the initial ALMA surveys of nearby star-forming regions.
    
    Finally, we would like to speculate on the possible time evolution of the
    \RvsLmm{} relation utilizing the three regions that have been observed at a
    similar sensitivity with ALMA, i.e. Lupus, Cha~I and USco. While \ChaI{}
    and \Lup{} have overlapping ages with large uncertainties (1-3 and
    2-3\,Myrs respectively), if we take these ages at face value, we see  a
    flattening of the \RvsLmm{} relation moving from the youngest to the oldest
    region.

\section{Summary} \label{sec_summary}

    Using ALMA archival data, we estimate the sizes of 152 protoplanetary
    disks in the three star-forming regions of \Lup{}, \ChaI{} and \USco{}.
    This results in the first homogenous estimation of dust disk sizes between these
    regions.  Because we use the same approach (visibility fitting using Nuker
    profiles) as in \cite{Tripathi2017} and \cite{Andrews2018a}, we add to our analysis
    their disk-size estimates for  \Oph{} and \Tau{}, for a total of 199 disk sizes from 5 different
    regions.  While the 5 regions have had their disk-sizes
    estimated in a consistent way, \Oph{} and \Tau{} were observed with the SMA, hence are biased towards the
    brightest millimeter disks.


    These 5 nearby regions cover the age range over which disks disperse (\OphFull{}:
    1-2 Myr; \TauFull{}: 1-3 Myr; \LupFull{}: 1-3 Myr; \ChaIFull{}:
    2-3 Myr; \UScoFull{}: 5-11 Myr) and host stars that span the entire stellar mass, 
    0.08 to 4.68 \Mstar{}; with typical (16-84\% quantile
    range) stellar masses of 0.17 to 0.89 \Mstar{}.


    Of the 199 disk-size estimates in our entire sample, there are 130 resolved
    disks and 69 unresolved disks for which we provide upper limits in the dust
    radii.  Estimated \R{68} disk sizes for the entire sample range from 8.5 to
    177\,au with a median of 39\,au.  For the 3 regions with \R{90}  estimates
    (\Lup{}, \ChaI{} and \USco{}), we find that \R{90} range from 10-231\,au
    with a median of 43\,au.

    Our main results can be summarized as follows:

\begin{itemize}

    \item For disks around near solar mass stars ($\Mstar{} \in
        [0.85-1.15\,M\odot$]) \R{90} disk sizes range from 20-103\,au.  This
        suggests that the $\sim45$\,au outer-edge of the Solar System, if primordial in origin, is not an
        outlier when compared with typical 1-3\,Myr disks, and that a
        truncating event (e.g. a stellar encounter or external
        photoevaporation) may not be required to explain its size.  Additionally we
        find that the mass of solids in the Solar System falls within the
        range of estimated dust-disk masses estimated for our sample.


    \item We find that the disks in \ChaI{} are not smaller than those in
        \Lup{} as previously suggested in \cite{Eisner2018}, whose comparison
        was based on a sub-sample of \Lup{} as well as samples from \Oph{} and \Tau{}
        all of which were biased towards higher luminosity sources.
        
    \item \USco{} disk sizes are not drawn from the same distribution of disk
        sizes as \Lup{} and \ChaI{}. In agreement with previous findings, the
        older \USco{} region appears to have a population of smaller disks,
        albeit not by a factor of 3 as reported in \cite{Barenfeld2017}.  We
        find that \USco{} disks are also smaller than those in \ChaI{}; our
        homogeneous analysis finds a difference in the median values of $\sim
        1.5$.

    \item Dust disk radii  correlate with millimeter luminosity in each of the
        5 regions. However, we find that the \RvsLmm{} relation is not the same
        in all regions but rather becomes flatter for older regions.
        Uncertainties in each region's age, lack of dynamic range in ages, and
        different stellar environments, impedes our ability to conclude that
        disk-size evolution with time is conclusively seen.
        
    \item We find no evidence for a correlation between the dust-disk outer
        radius and stellar mass in any of the 5 regions we tested.

    \item Only in the Lupus star-forming region is there a modest correlation
        between the dust outer radius and the stellar luminosity.

\end{itemize}

   In relation to the last two points, one should keep in mind that we are
   examining only a subset of the entire disk population in each region.  While
   we find a  correlation between \Lmm{} and \Mstar{} for our sub-samples, in
   each region the correlation is  shallower than when the entire population is
   included (see Section \ref{sec_modeling_sample_selection} and Appendix
   \ref{app_LmmvsMstar}). Therefore, we can conclude that dust radii correlate
   less with \Mstar{} and \Lstar{} than the millimeter luminosity. Deeper ALMA
   observations, especially of disks around low-mass stars, will be necessary
   to test if a modest correlation is present between dust disk sizes and
   stellar properties. \\

\vspace{2cm}
\acknowledgments

N. H. and I.P. acknowledge support from a NSF Astronomy \& Astrophysics
Research Grant (ID: 1515392). The results reported herein benefitted from
collaborations and/or information exchange with the {\it Earths in Other Solar
Systems} team which is part of NASA’s Nexus for Exoplanet System Science
(NExSS) research coordination network sponsored by NASA’s Science Mission
Directorate.
P.P. acknowledges support provided by the Alexander von Humboldt
Foundation in the framework of the Sofja Kovalevskaja Award endowed by the
Federal Ministry of Education and Research.
M.T. has been supported by the UK Science and Technology research Council
(STFC), and by the European Union’s Horizon 2020 research and innovation
programme under the Marie Sklodowska-Curie grant agreement No. 823823 (RISE
DUSTBUSTERS project).
R.M. acknowledges research support from NSF grant AST-1824869 and NASA-XRP grant 80NSSC18K0397.
This paper makes use of the following ALMA data:
ADS/JAO.ALMA\#
2011.0.00526.S,
2013.1.00220.S, 
2013.1.00395.S,
2013.1.00437.S.
ALMA is a partnership of ESO (representing its member states), NSF (USA) and
NINS (Japan), together with NRC (Canada), MOST and ASIAA (Taiwan), and KASI
(Republic of Korea), in cooperation with the Republic of Chile. The Joint ALMA
Observatory is operated by ESO, AUI/NRAO and NAOJ.  The National Radio
Astronomy Observatory is a facility of the National Science Foundation operated
under cooperative agreement by Associated Universities, Inc.
Thanks the LSSTC Data Science Fellowship Program, which is funded by LSSTC, NSF
Cybertraining Grant \#1829740, the Brinson Foundation, and the Moore Foundation;
participation in the program by N. H. has benefited this work.  Finally, the
authors would like to thank Tom Zega and Pierre Haenecour for useful
discussions that aided this work.

%

\vspace{5mm}
\facilities{ALMA, GAIA}


\software{scipy, astropy \citep{AstropyCollaboration2013,AstropyCollaboration2018}, GALARIO \citep{Tazzari2018}, emcee \citep{Foreman-Mackey2013}, linmix \citep[][https://github.com/jmeyers314/linmix]{Kelly2007}, uvplot \citep{Tazzariuvplot}}



\appendix

\input{appendix-comparison_with_literature_results.tex}

\input{appendix-stellar_properties.tex}

\input{appendix-sample_selection.tex}

\input{appendix-LmmvsMstar_relation.tex}

\input{appendix-correlation_r90_r68.tex}

\input{appendix-individual_modeling_results.tex}

\input{appendix-LmmvsLstar_relation.tex}





\FloatBarrier


\bibliographystyle{aasjournal}
\bibliography{ads_auto,custom,solar_system_references,Hendler2017}



\end{document}

%% file: appendix-comparison_with_literature_results.tex
\section{Comparison with literature results}

\subsection{Lupus: comparing \cite{Andrews2018a} with this work} \label{app_lupus_comparison}

\cite{Tripathi2017} and \cite{Andrews2018a} have used the same modeling approach
to infer disk sizes and their papers combined provide the largest compilation
available in the literature. To further expand upon their work, we have applied
the same steps to model the disks in Cha~I and USco. However, there are bound
to be small differences in chosen modeling and fitting parameters and
techniques, as well as differences in determining uncertainties and
upper-limits, hence our re-analysis of the Lupus dataset and the
comparison presented here.

\begin{figure*}[ht!]
    \plotone{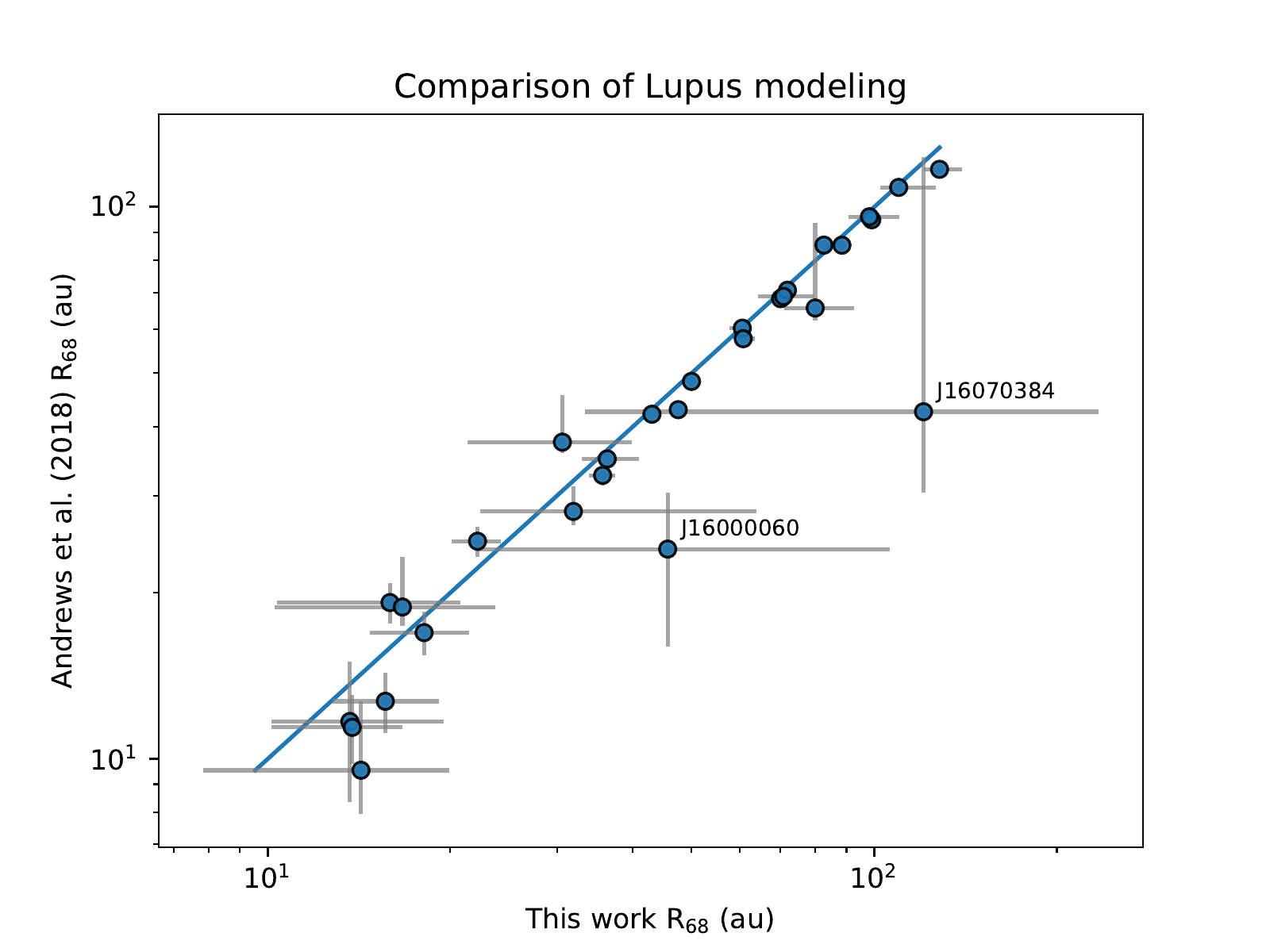}
    \caption{A comparison of disk sizes calculated in this work with sizes from \cite{Andrews2018a}.  The 1:1 line is shown in blue.\label{fig_andrews_vs_me}}
\end{figure*}

Figure \ref{fig_andrews_vs_me} shows our Lupus disk sizes vs those reported in  \cite{Andrews2018a}.  For most models there is tight agreement in disk sizes between the two works and  for almost all sources there is agreement within the uncertainties. 

\begin{figure*}[ht!]
    \plottwo{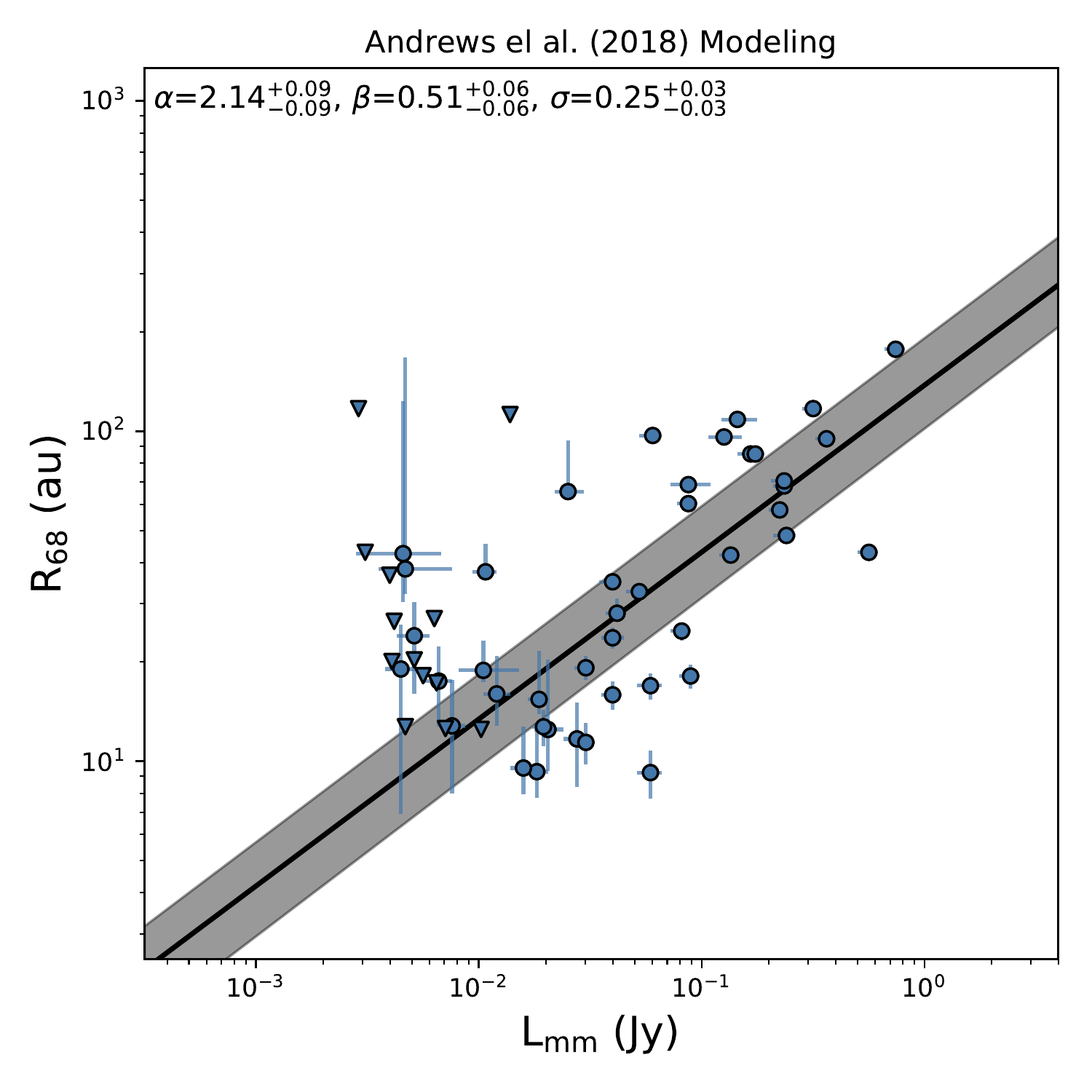}{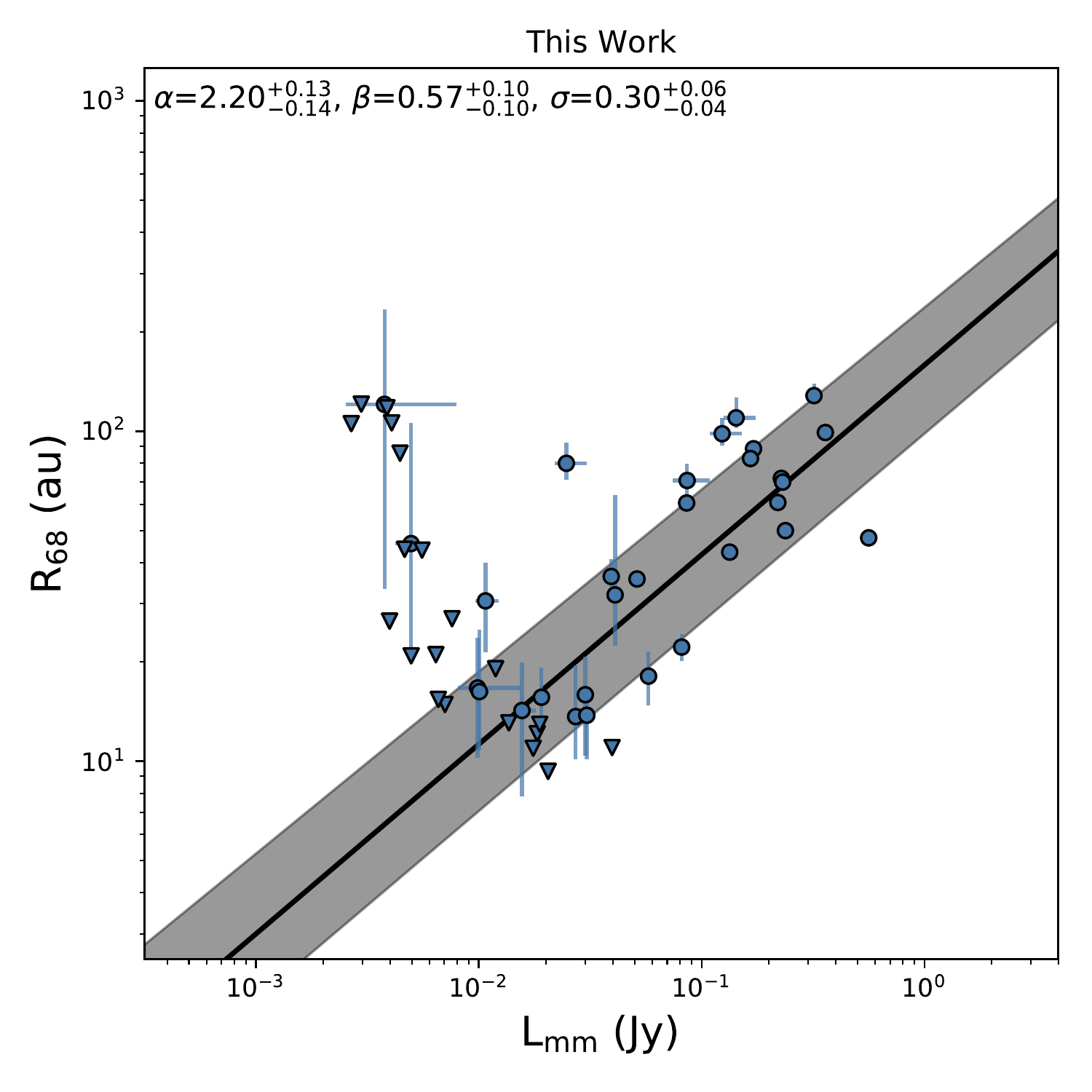}
    \caption{Two panels showing the results of Bayesian linear regression fitting using disk sizes from \cite{Andrews2018a} (left) and this work (right).  The upper-left corner of each panel includes
    the best fit parameters ($\alpha$ and $\beta$) and scatter of the relation ($\sigma$), see eq.~2 in the main text. \label{fig_Lmm_andrews_vs_me}}
\end{figure*}

One caveat about Figure \ref{fig_andrews_vs_me} is that it does not compare
disks for which only an upper limit to the disk size could be estimated. In general, our method results in larger uncertainties than those given in \cite{Andrews2018a}. For this
reason our upper limits are typically larger, and consequently for 10 disks
which \cite{Andrews2018a} report a size for, we  only provide an upper limit. Figure \ref{fig_Lmm_andrews_vs_me} presents the results of Bayesian linear
regression fitting \citep{Kelly2007} of both modeling results with these
differences in uncertainties and upper limits included.  Our fitting of both
the \cite{Andrews2018a} sample and our sample
gives results that are consistent with each other, see values in
Figure \ref{fig_Lmm_andrews_vs_me} and in Table~1 of \cite{Andrews2018a}.

\subsection{High- vs medium-resolution observations} \label{app_high_vs_low}

Due to the long exposure times required to make observations of disks at $\le
0.1''$ like those in \cite{Long2018} and DSHARP \citep{Andrews2018}, it is
unlikely that large sample sizes required to compare disk demographics of
star-forming regions will be available in the near-term at such high
resolution.  This means that medium-resolution observations, like those
analyzed in this work, have an important role to play in establishing what are
typical disk properties, e.g. disk sizes, and their spread.  However, that can
only be the case if these observations truly inform us about the disk size.
Here, we test the agreement between disk sizes derived from high- and
medium-resolution observations.

There are 26 \Tau{} sources that are common between the high-resolution
observations of \cite{Long2018}, \cite{Huang2018} and \cite{Long2019}
and the medium resolution observations of \cite{Tripathi2017}.  The
medium-resolution disk sizes used here are from \cite{Andrews2018a} who updated
the \cite{Tripathi2017} results by including GAIA DR2 distances.  We use Eq.
\ref{eq_r90_r68} (see Appendix \ref{app_correlation_r90_r68}) to convert the \cite{Tripathi2017} disk sizes from \R{68} to
the \R{90} radii.  It should be noted that the high-resolution disk sizes
include a mix of \R{95} and \R{90} emission radii but these values are so close
to each other that we do not attempt to derive an additional scaling.

\begin{figure}
    \plotone{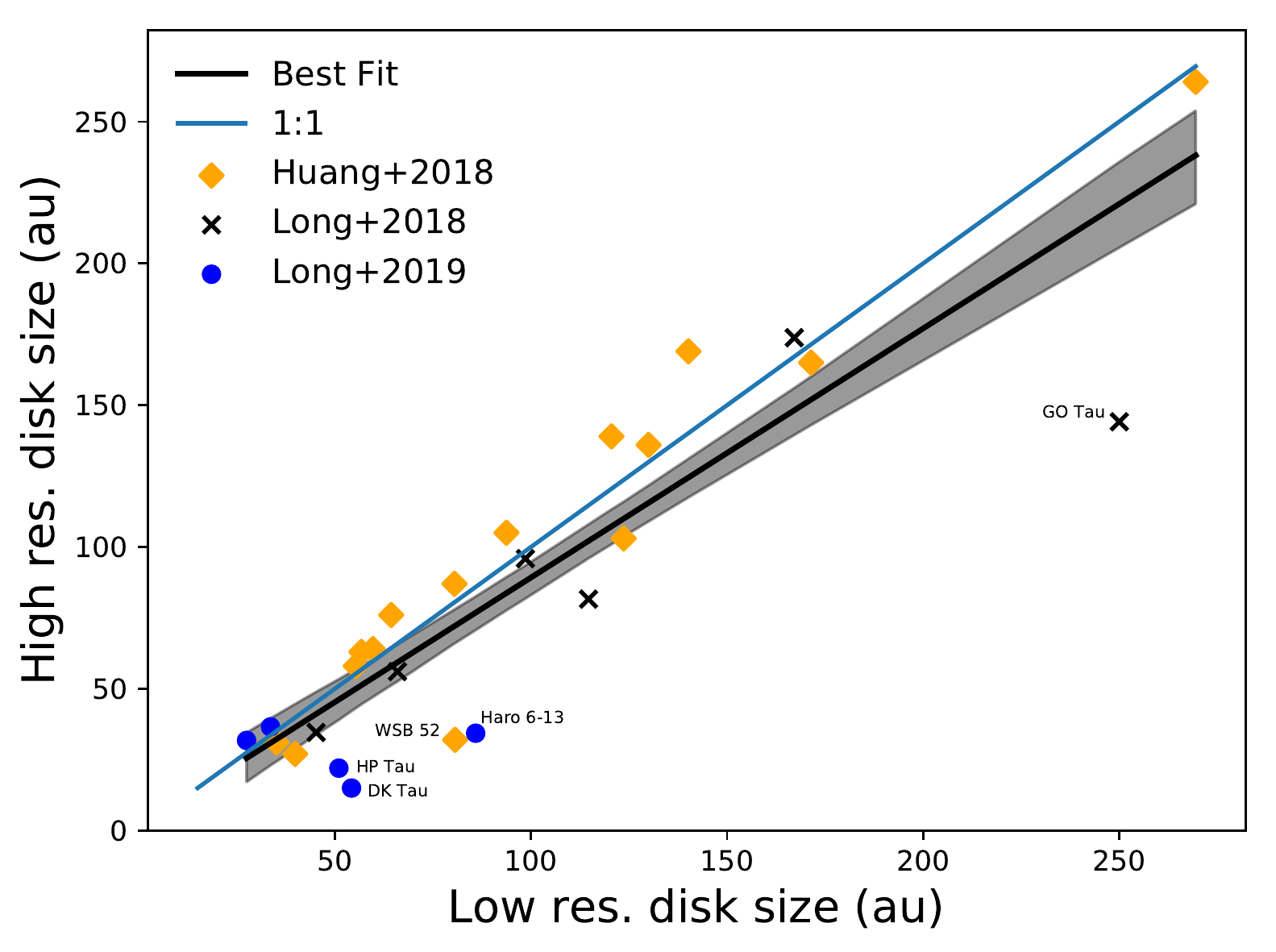}
    \caption{A comparison of disk size measurements for \Tau{} disks that have been resolved at both high-resolution (ALMA) and medium-resolution (SMA).  The medium resolution observations are taken from \cite{Tripathi2017}.  High-resolution observations are from \cite{Huang2018}, \cite{Long2018} and \cite{Long2019}
    (indicated by a diamond, x and circle respectively).  For each disk size measurement, the reported uncertainties are all 1~au or less. The best fit from Bayesian linear regression fitting is shown in black, while the corresponding $1\sigma$ confidence interval is shown in grey.  A 1:1 relation line is shown in blue.\label{fig_high_vs_low}}
\end{figure}

We directly compare the high- and medium-resolution disk size data in Figure
\ref{fig_high_vs_low} along with a 1:1 line (in blue) for reference.  The
relationship between the high-resolution (\R{\text{high}}) and low-resolution
(\R{\text{low}}) disk sizes is fit using Bayesian linear regression fitting
\citep{Kelly2007} and the best-fit is shown in black with the corresponding
$1\sigma$ confidence interval in gray.  The relationship is also given as Eq.
\ref{eq_highres_lowres}.

\begin{equation} \label{eq_highres_lowres}
    \R{\text{high}}(\R{\text{low}}) = 0.95 (\text{au}) + 0.88 \R{\text{low}} (\text{au})
\end{equation}

%% file: appendix-stellar_properties.tex
\section{Stellar and Disk Properties}
\label{app_properties}

Table \ref{tab_stellar} lists stellar properties and derived stellar masses ($\Mstar$) as described in Section \ref{sec_stellar_properties}.
To derive stellar masses we follow the approach used in \cite{Pascucci2016} and assume an uncertainty of
0.02\,dex in the stellar temperature for spectral types earlier than M3 and
0.01\,dex for later spectral types, and a 0.1\,dex uncertainty on all stellar
luminosities.

\input{table-stellar_properties.tex}

%% file: table-stellar_properties.tex
\startlongtable
\begin{deluxetable*}{lllllll}

    \tablecaption{Stellar and Disk Properties\label{tab_stellar}}
    \tablewidth{0pt}
    \tablehead{
\colhead{2MASS} &
\colhead{Region} &
\colhead{Distance} &
\colhead{$\log(\Lstar)$} &
\colhead{T$_{\text{eff}}$} &
\colhead{$\log(\Mstar)$} &
\colhead{L$_{\text{mm}}$\tablenotemark{a}}
\\
\colhead{} &
\colhead{} &
\colhead{(pc)} &
\colhead{($\Lsun$)} &
\colhead{(K)} &
\colhead{($M_\odot$)} &
\colhead{(mJy)}
}
\startdata
J10533978-7712338&ChaI&$190.88^{+3.85}_{-3.70}$&$-1.70^{+0.01}_{-0.01}$&$3560.00$&$-0.48^{+0.10}_{-0.06}$&$6.95^{+0.59}_{-0.55}$\\ 
J10555973-7724399&ChaI&$184.11^{+1.07}_{-1.06}$&$-0.74^{+0.11}_{-0.11}$&$4060.00$&$-0.11^{+0.14}_{-0.10}$&$58.97^{+3.00}_{-2.93}$\\ 
J10561638-7630530&ChaI&$195.52^{+4.15}_{-3.98}$&$-1.10^{+0.05}_{-0.05}$&$2935.00$&$-0.89^{+0.02}_{-0.02}$&$7.78^{+0.66}_{-0.61}$\\ 
J10563044-7711393&ChaI&$182.12^{+0.65}_{-0.64}$&$-0.37^{+0.25}_{-0.25}$&$4060.00$&$-0.14^{+0.19}_{-0.15}$&$198.98^{+3.30}_{-3.25}$\\ 
J10580597-7711501&ChaI&$185.69^{+3.70}_{-3.56}$&$-2.00^{+0.01}_{-0.01}$&$3060.00$&$-0.96^{+0.03}_{-0.02}$&$4.71^{+0.48}_{-0.45}$\\ 
J10581677-7717170&ChaI&$188.83^{+1.50}_{-1.48}$&$0.28^{+1.11}_{-1.11}$&$5250.00$&$0.21^{+0.16}_{-0.14}$&$564.26^{+10.87}_{-10.60}$\\ 
J10590108-7722407&ChaI&$184.23^{+0.65}_{-0.65}$&$-0.42^{+0.22}_{-0.22}$&$4060.00$&$-0.13^{+0.17}_{-0.15}$&$113.14^{+3.77}_{-3.72}$\\ 
J10590699-7701404&ChaI&$186.47^{+0.80}_{-0.80}$&$0.51^{+1.91}_{-1.91}$&$5110.00$&$0.27^{+0.33}_{-0.24}$&$784.48^{+8.14}_{-8.03}$\\
\enddata
\tablenotetext{a}{For consistency, our definition of \Lmm{} is the same as found in \cite{Andrews2018a}.  It is a flux density scaled to 140 pc.}
\end{deluxetable*}

%% file: appendix-sample_selection.tex
\section{Determination and removal of spurious modeling}
\label{app_sample_selection} 
\begin{figure}[ht!]
    \plotone{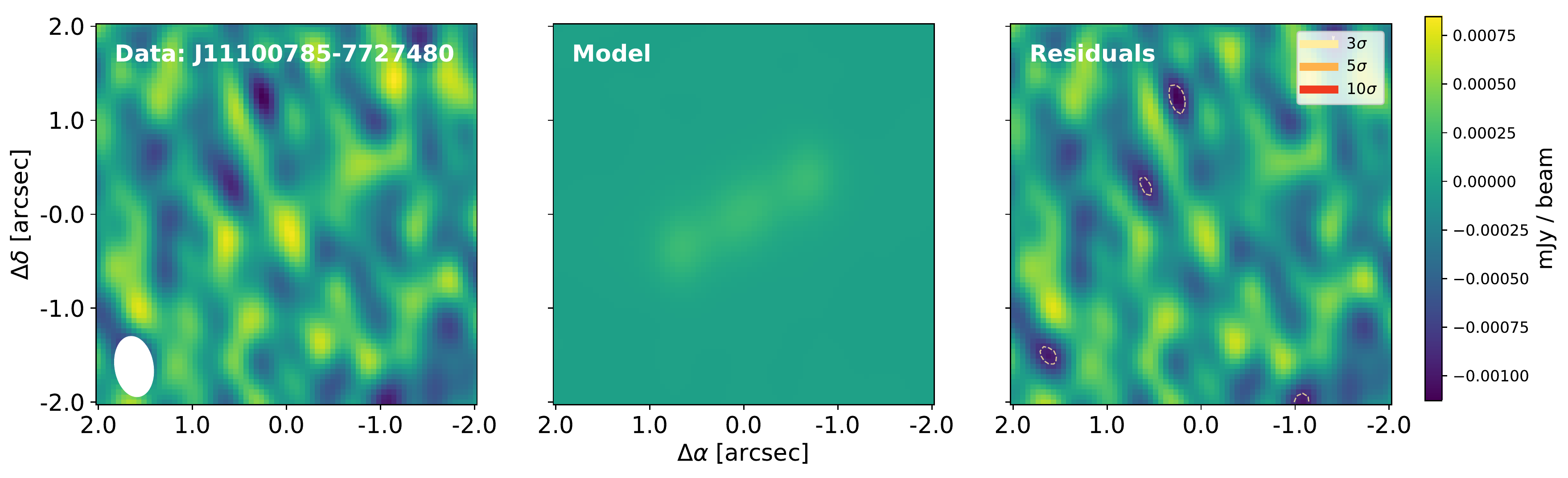}
    \caption{Example of a Nuker profile fitting the noise and not the faint disk emission from J11100785-7727480. The flux density is only $0.49\pm0.16$ mJy \citep{Pascucci2016}, a marginal 3$\sigma$ detection.
    \label{fig_poor_modeling_residuals}}
\end{figure}

During our model fitting we realized that low
signal-to-noise data can result in Nuker profiles that fit the noise with a
disk model that is typically extended, faint, highly inclined and/or ring-like.
An example is shown in Figure~\ref{fig_poor_modeling_residuals}.  In these
cases the uncertainties on the disk sizes are large.  In order to cull these
models from our analysis, we examine by eye sources where the uncertainty in
disk size is larger than 66\% of the disk size.  By observing the fitting in
the UV plane, and comparing continuum images of the data and model, we can
identify these cases.  Of the 23 disks that meet this metric, we find 15 that
are obviously problematic, and hence removed them from our analysis (see
Section~\ref{sec_modeling_sample_selection}).

Sources removed from the \ChaI{} sample are: J11100785-7727480, J11064180-7635489, J11092266-7634320, J11071206-7632232 and J11045701-7715569.

Sources removed from the \USco{} sample are: J16095933-1800090, J16095361-1754474, J16115091-2012098, J16102857-1904469, J16073939-1917472, J15534211-2049282, J16035793-1942108, J16043916-1942459, J16135434-2320342 and J16303390-2428062

%% file: appendix-LmmvsMstar_relation.tex
\section{\LmmvsMstar{} Relation}
\label{app_LmmvsMstar}

Our sample population is a subset of the entire disk population in each region,
and is biased towards resolved sources.  In order to understand this bias, we
test our sample for a correlation between \Lmm{} and \Mstar{} so that we can compare
it to previously measured \Lmm{}--\Mstar{} correlations
\citep[e.g.][]{Ansdell2016,Pascucci2016}.  We find a shallower slope for the
\LmmvsMstar{} correlation, and discuss this in detail at the end of Section
\ref{sec_modeling}.

\begin{figure*}
    \plotone{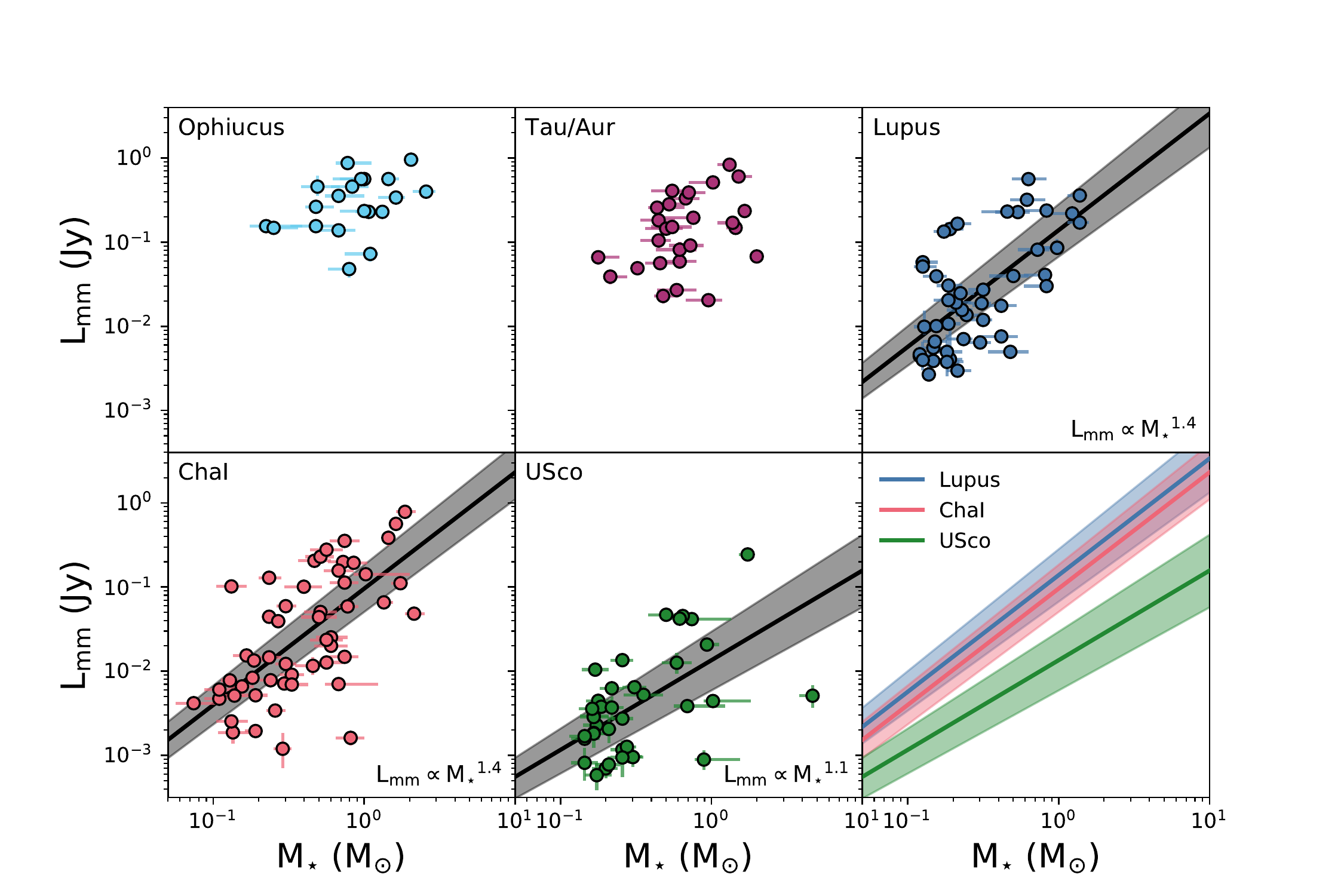}
    \label{fig_Lmm_vs_Mstar}
    \caption{Fitting of \LmmvsMstar{}. The first 5 panels (left to right; top to bottom; ordered by region age) show the model results of each region as circles (resolved) and triangles (upper-limits).  The best fit from MCMC linear regression is plotted as a black line, and surrounded by our 68\% confidence intervals in grey.  The last panel replots the bests fits of each region (and the corresponding 68\% confidence intervals) so that they can be directly compared. Fit parameters for each region are given in Table \ref{tab_Lmm_vs_Mstar}.}
\end{figure*}

\begin{deluxetable*}{@{\extracolsep{4pt}}lllllllllllll}[ht]
\tablecaption{\LmmvsMstar{} statistical tests \label{tab_Lmm_vs_Mstar}}
\tablewidth{0pt}
\tablehead{
\colhead{} &
\multicolumn2c{Shapiro \Lmm{}} &
\multicolumn2c{Shapiro \Mstar{}} &
\multicolumn2c{Pearson r} &
\multicolumn2c{Spearman $\rho$} & 
\multicolumn4c{Regression Parameters} \\
\cline{2-3}
\cline{4-5}
\cline{6-7}
\cline{8-9}
\cline{10-13}
\colhead{Region} &
\colhead{stat} &
\colhead{p-value} &
\colhead{stat} &
\colhead{p-value} &
\colhead{stat} &
\colhead{p-value} &
\colhead{stat} &
\colhead{p-value} &
\colhead{$\alpha$}  & 
\colhead{$\beta$}  & 
\colhead{$\sigma$} &
\colhead{$\hat{\rho}$}
}
\startdata
    \Oph{}  & 0.91 & 6.71e-02 & 0.91 & 5.66e-02 & 0.38       & 9.86e-02       & \dem{0.30} & \dem{2.01e-01} & \dem{$-0.53^{+0.083}_{-0.08}$} & \dem{$0.49^{+0.36}_{-0.36}$} & \dem{$0.35^{+0.074}_{-0.056}$} & $0.61^{+0.17}_{-0.28}$\\
    \Tau{}  & 0.87 & 3.25e-03 & 0.82 & 2.86e-04 & \dem{0.37} & \dem{6.06e-02} & 0.37       & 5.54e-02       & \dem{$-0.77^{+0.1}_{-0.1}$   } & \dem{$0.75^{+0.36}_{-0.36}$} & \dem{$0.43^{+0.074}_{-0.058}$} & $0.66^{+0.12}_{-0.18}$\\
    \Lup{}  & 0.77 & 2.89e-07 & 0.67 & 3.46e-09 & \dem{0.56} & \dem{3.82e-05} & 0.61       & 3.94e-06       & $-0.86^{+0.2}_{-0.1}$ & $1.4^{+0.2}_{-0.2}$ & $0.51^{+0.06}_{-0.05}$ & $0.81^{+0.05}_{-0.06}$\\
    \ChaI{} & 0.8  & 1.19e-07 & 0.6  & 1.35e-11 & \dem{0.61} & \dem{2.78e-07} & 0.55       & 6.20e-06       & $ -1^{+0.1}_{-0.1}$   & $1.4^{+0.2}_{-0.2}$ & $0.56^{+0.06}_{-0.06}$ & $0.82^{+0.05}_{-0.06}$\\
    \USco{} & 0.45 & 8.04e-11 & 0.34 & 7.70e-12 & \dem{0.29} & \dem{7.48e-02} & 0.49       & 1.81e-03       & $-1.9^{+0.2}_{-0.2}$  & $1.1^{+0.3}_{-0.3}$ & $0.52^{+0.07}_{-0.06}$ & $0.75^{+0.08}_{-0.1}$\\
\enddata
    \tablecomments{Values we consider unreliable are greyed out (see Section~\ref{sec_relations}).}
\end{deluxetable*}

%% file: appendix-correlation_r90_r68.tex

\section{Correlation of \R{90} and \R{68}} \label{app_correlation_r90_r68}

Different emission fractions have been used to define the location of \R{eff}
in past works.  For example, \cite{Tripathi2017} and \cite{Andrews2018a}, who analyze low- and medium-resolution SMA and ALMA data, calculate the radius containing 68\% of the total flux (\R{68}). However, high-resolution ALMA images have shown that \R{68} does not properly capture the radial extent of the disk, several sub-structures are left out, hence more recent works compute and analyze the 90\% (\R{90}) or 95\% (\R{95}) dust radius (see \citealt{Long2018} and \citealt{Huang2018}).   In order to
compare the results of this work with disk sizes from the literature, we are interested in understanding if \R{68} is a reliable predictor of \R{90}.

\begin{figure}[ht!]
    \plotone{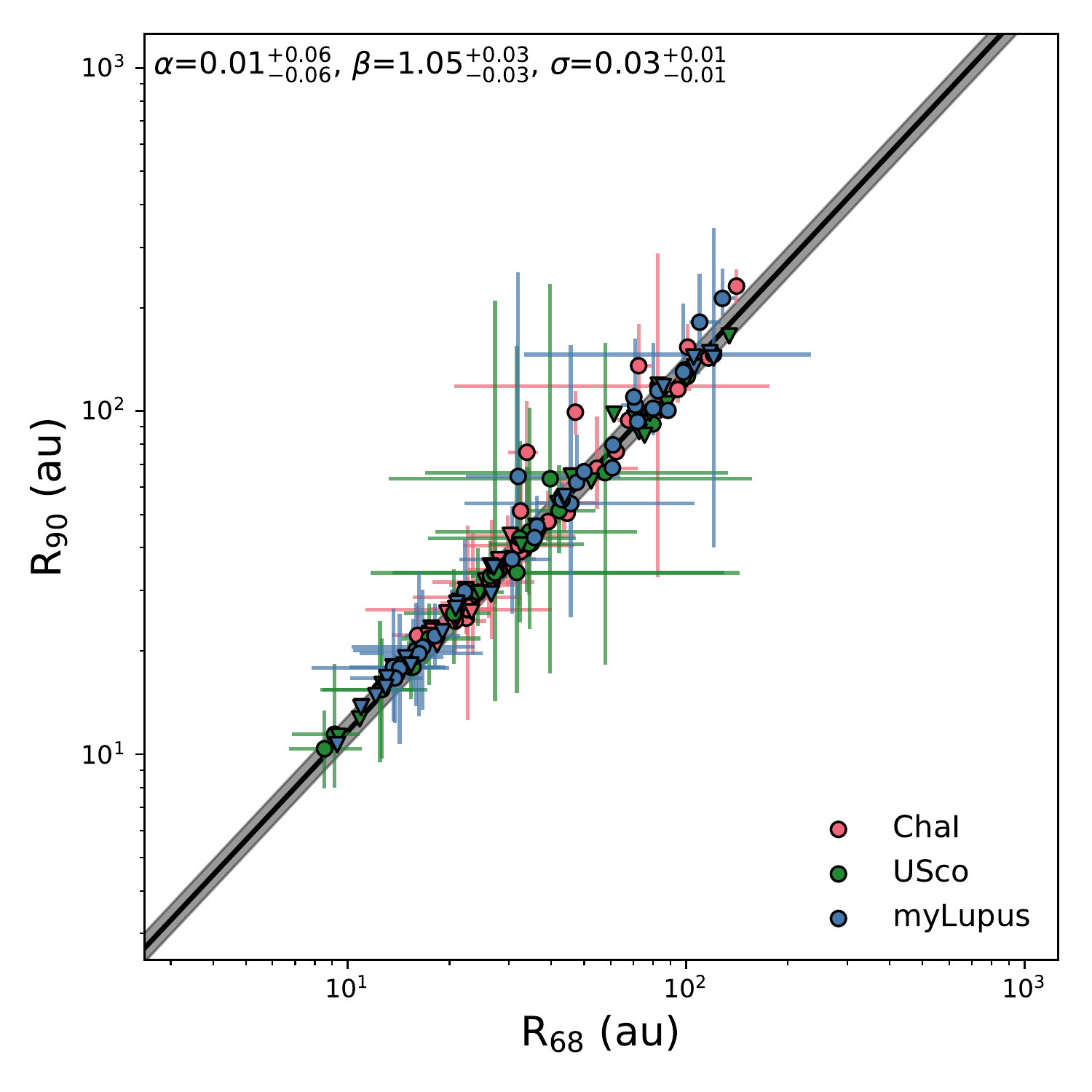}
    \caption{Here we compare our \R{68} dust-disk size estimates with the \R{90} size for each disk.  Constrained disks sizes are plotted as circles while upper-limits are given as triangles.  Symbols are color coded according to their associated region.  The correlation between \R{68} and \R{90} is measured by fitting the values using linmix.  The best fit is shown as a black line, with a 1-sigma confidence interval in gray.  The parameters resulting from the mcmc chain are given in the upper left of the figure (see Equation \ref{eq_r90_r68}). \label{fig_rout_R90_vs_R68_N1}}
\end{figure}

The 68\% and 90\% emission radii we measure from modeled disks in
\Lup{}, \ChaI{}, and \USco{} are shown in Figure \ref{fig_rout_R90_vs_R68_N1}.
Visual inspection suggests that the two disk sizes are strongly correlated.
We calculate the Spearman $\rho$ rank-order correlation coefficient\footnote{We cannot apply the Pearson correlation test, which tests for a linear relationship between two quantities, because we find that \R{68} and \R{90} are not bivariate normally distributed.} to test the null hypothesis that the two variables are uncorrelated. As shown in Table
\ref{tab_rout_R90_vs_R68_N1}, the  correlation coefficient is nearly unity (positive rank order) with an extremely low probability (p-value) that \R{68} and \R{90} are uncorrelated.

We then fit the data using the Bayesian linear regression method developed by \cite{Kelly2007}
and report the
results in Table \ref{tab_rout_R90_vs_R68_N1}.  
 Using these best
fitting parameters, we can convert the \R{68} to a \R{90} disk size:.

\begin{deluxetable*}{@{\extracolsep{4pt}}lllllllllll}
\tablecaption{\R{90} vs \R{68} statistical analysis and fitting \label{tab_rout_R90_vs_R68_N1}}
\tablewidth{0pt}
\tablehead{
    \colhead{} &
    \multicolumn2c{Shapiro \R{68}} &
    \multicolumn2c{Shapiro \R{90}} &
    \multicolumn2c{Spearman $\rho$} &
    \multicolumn4c{Regression Parameters}
    \\
    \cline{2-3}
    \cline{4-5}
    \cline{6-7}
    \cline{8-11}
    \colhead{} &
    \colhead{stat} &
    \colhead{p-value} &
    \colhead{stat} &
    \colhead{p-value} &
    \colhead{stat} &
    \colhead{p-value} &
    \colhead{$\alpha$}  & 
    \colhead{$\beta$}  & 
    \colhead{$\sigma$} &
    \colhead{corr}
}
\startdata
    \R{90} vs \R{68} & 0.87  & 3.1e-07  & 0.84 & 4.1e-08  & 0.98  & 2.6e-63  & 0.01$^{+0.06}_{-0.06}$  & 1.05$^{+0.03}_{-0.03}$  & 0.03$^{+0.01}_{-0.01}$ & $1.00^{+0.00}_{-0.00}$\\
\enddata
\end{deluxetable*}

\begin{equation} \label{eq_r90_r68}
    \log{\R{90}} = -0.037 \text{(au)} + 1.1 \log{\R{68}} \text{(au)}
\end{equation}

%% file: appendix-individual_modeling_results.tex
\section{Individual modeling results}
\label{app_individual_modeling_results}

\input{table-modeling_results.tex}

%% file: table-modeling_results.tex
\begin{longrotatetable}
\begin{deluxetable}{lllllllllllll}

    \tablecaption{Modeling Parameters\label{tab_modeling}}
    \tablewidth{0pt}
    \tablehead{
\colhead{2MASS} &
\colhead{$R_{68}$} &
\colhead{$R_{90}$} &
\colhead{$R_{t}$} &
\colhead{$f_{0}$}&
\colhead{$\alpha$} &
\colhead{$\beta$} &
\colhead{$\gamma$} &
\colhead{$i$} &
\colhead{PA} &
\colhead{dDec} &
\colhead{dRA} &
\colhead{lnwcorr} 
\\
\colhead{} &
\colhead{$\arcsec$} &
\colhead{$\arcsec$} &
\colhead{$\arcsec$} &
    \colhead{(Jy/Sr)}&
\colhead{} &
\colhead{} &
\colhead{} &
    \colhead{($^\circ$)} &
    \colhead{($^\circ$)} &
    \colhead{($\arcsec$)} &
    \colhead{($\arcsec$)} &
\colhead{}
}
\startdata
J10533978-7712338&$<0.11$&$<0.13$&$0.11^{+0.05}_{-0.05}$&$9.98^{+0.52}_{-0.35}$&$60.00^{+32.12}_{-40.13}$&$9.37^{+3.55}_{-3.31}$&$-2.03^{+1.69}_{-2.69}$&$51.64^{+24.41}_{-28.56}$&$58.51^{+58.02}_{-37.85}$&$-0.01^{+0.03}_{-0.03}$&$-0.45^{+0.03}_{-0.03}$&$-0.96^{+0.00}_{-0.00}$\\ 
J10555973-7724399&$0.11^{+0.03}_{-0.01}$&$0.13^{+0.03}_{-0.03}$&$0.10^{+0.02}_{-0.02}$&$10.84^{+0.44}_{-0.16}$&$52.10^{+36.53}_{-26.66}$&$8.68^{+3.24}_{-3.09}$&$-2.75^{+2.09}_{-2.20}$&$43.74^{+11.77}_{-13.83}$&$88.60^{+29.78}_{-26.10}$&$-0.07^{+0.01}_{-0.01}$&$-0.07^{+0.01}_{-0.01}$&$-3.79^{+0.01}_{-0.01}$\\ 
J10561638-7630530&$<0.12$&$<0.14$&$0.10^{+0.04}_{-0.04}$&$10.23^{+0.49}_{-0.51}$&$56.36^{+35.09}_{-41.50}$&$9.79^{+3.28}_{-4.37}$&$-2.55^{+1.85}_{-1.57}$&$41.79^{+25.26}_{-17.18}$&$62.34^{+25.32}_{-23.98}$&$-0.07^{+0.02}_{-0.03}$&$-0.37^{+0.02}_{-0.01}$&$-0.49^{+0.01}_{-0.00}$\\ 
J10563044-7711393&$0.77^{+0.03}_{-0.03}$&$1.27^{+0.15}_{-0.15}$&$0.54^{+0.05}_{-0.04}$&$10.01^{+0.05}_{-0.10}$&$65.43^{+31.18}_{-46.05}$&$4.27^{+0.44}_{-0.36}$&$-3.57^{+1.37}_{-1.73}$&$51.64^{+1.27}_{-1.09}$&$167.35^{+1.46}_{-1.92}$&$0.06^{+0.01}_{-0.01}$&$-0.41^{+0.01}_{-0.01}$&$-3.88^{+0.01}_{-0.01}$\\ 
J10580597-7711501&$<0.11$&$<0.13$&$0.11^{+0.03}_{-0.05}$&$10.01^{+0.47}_{-0.70}$&$62.61^{+28.85}_{-29.76}$&$9.59^{+2.91}_{-3.74}$&$-2.06^{+1.74}_{-2.43}$&$55.33^{+18.43}_{-23.39}$&$147.06^{+17.04}_{-29.60}$&$-0.02^{+0.03}_{-0.03}$&$-0.37^{+0.03}_{-0.03}$&$-0.48^{+0.00}_{-0.00}$\\ 
J10581677-7717170&$0.62^{+0.01}_{-0.01}$&$0.76^{+0.03}_{-0.02}$&$0.57^{+0.02}_{-0.03}$&$10.43^{+0.06}_{-0.03}$&$65.83^{+27.95}_{-24.66}$&$7.60^{+1.27}_{-0.75}$&$-3.74^{+0.49}_{-0.86}$&$42.01^{+0.98}_{-0.77}$&$157.12^{+1.36}_{-1.56}$&$-0.06^{+0.00}_{-0.00}$&$-0.38^{+0.00}_{-0.00}$&$-4.10^{+0.01}_{-0.01}$\\
\enddata
\end{deluxetable}
\end{longrotatetable}

%% file: appendix-LmmvsLstar_relation.tex
\section{\LmmvsLstar{} Relation}
\label{app_LmmvsLstar}

In our discussion on disk sizes scaling relations (see Section \ref{sec_discussion_scaling_relations}), we reproduce the temperature power law index ($q$) estimation following the method given in \cite{Andrews2018a}.  This approach requires us to know the \LmmvsLstar for our samples, which we provide here using the statistical methods described in Section \ref{sec_results}.

\begin{figure*}
    \plotone{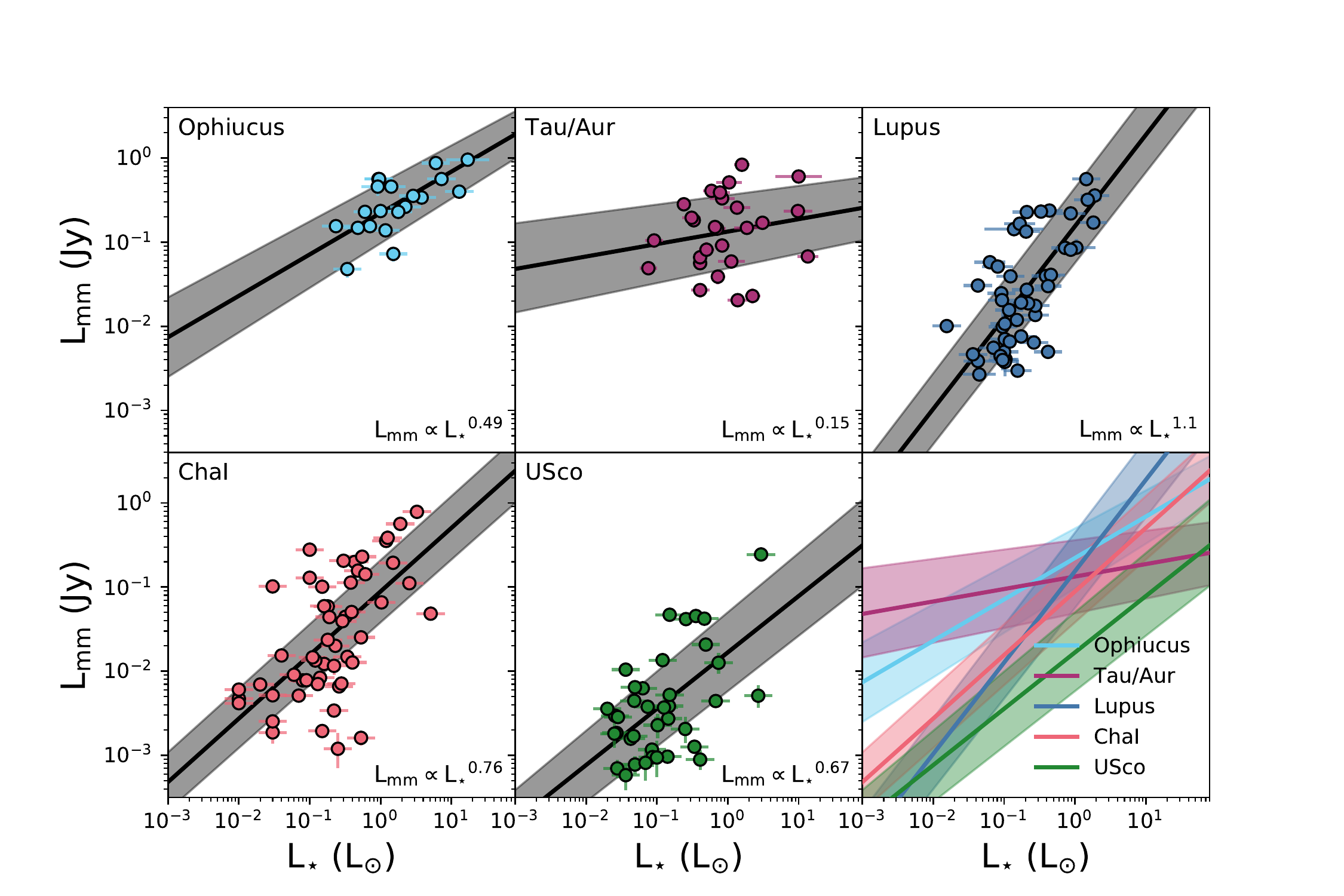}
    \label{fig_Lmm_vs_Lstar}
    \caption{Fitting of \LmmvsLstar{}. The first 5 panels (left to right; top to bottom; ordered by region age) show the model results of each region as circles (resolved) and triangles (upper-limits).  The best fit from MCMC linear regression is plotted as a black line, and surrounded by our 68\% confidence intervals in grey.  The last panel replots the bests fits of each region (and the corresponding 68\% confidence intervals) so that they can be directly compared. Fit parameters for each region are given in Table \ref{fig_Lmm_vs_Lstar}.}
\end{figure*}

\begin{deluxetable*}{@{\extracolsep{4pt}}lllllllllllll}[ht]
\tablecaption{\LmmvsLstar{} statistical tests \label{tab_Lmm_vs_Lstar}}
\tablewidth{0pt}
\tablehead{
\colhead{} &
\multicolumn2c{Shapiro \Lmm{}} &
\multicolumn2c{Shapiro \Lstar{}} &
\multicolumn2c{Pearson r} &
\multicolumn2c{Spearman $\rho$} & 
\multicolumn4c{Regression Parameters} \\
\cline{2-3}
\cline{4-5}
\cline{6-7}
\cline{8-9}
\cline{10-13}
\colhead{Region} &
\colhead{stat} &
\colhead{p-value} &
\colhead{stat} &
\colhead{p-value} &
\colhead{stat} &
\colhead{p-value} &
\colhead{stat} &
\colhead{p-value} &
\colhead{$\alpha$}  & 
\colhead{$\beta$}  & 
\colhead{$\sigma$} &
\colhead{$\hat{\rho}$}
}
\startdata
    \Oph{}  & 0.66 & 1.3e-05 & 0.90 & 5.0e-02 & \dem{0.66} & \dem{1.6e-03} & 0.57 & 8.7e-03 & $-0.65^{+0.07}_{-0.07}$ & $0.49^{+0.2}_{-0.2}$ & $0.27^{+0.07}_{-0.05}$ & $0.84^{+0.08}_{-0.1}$\\
    \Tau{}  & 0.55 & 6.2e-08 & 0.82 & 3.5e-04 & \dem{0.15} & \dem{4.6e-01} & 0.16 & 4.2e-01 & $-0.88^{+0.09}_{-0.09}$ & $0.15^{+0.2}_{-0.2}$ & $0.46^{+0.08}_{-0.06}$ & $0.42^{+0.2}_{-0.2}$\\
    \Lup{}  & 0.66 & 3.0e-09 & 0.67 & 3.5e-09 & \dem{0.73} & \dem{3.4e-09} & 0.63 & 1.7e-06 & $-0.81^{+0.1}_{-0.1}$   & $1.1^{+0.2}_{-0.2}$  & $0.47^{+0.06}_{-0.06}$ & $0.86^{+0.04}_{-0.06}$\\
    \ChaI{} & 0.54 & 7.3e-12 & 0.62 & 1.1e-10 & \dem{0.52} & \dem{4.5e-05} & 0.54 & 1.8e-05 & $-1.1^{+0.1}_{-0.1}$    & $0.76^{+0.1}_{-0.1}$ & $0.59^{+0.07}_{-0.06}$ & $0.79^{+0.06}_{-0.07}$\\
    \USco{} & 0.45 & 1.0e-10 & 0.34 & 7.7e-12 & \dem{0.69} & \dem{1.6e-06} & 0.46 & 3.9e-03 & $-1.8^{+0.2}_{-0.2}$    & $0.67^{+0.2}_{-0.2}$ & $0.53^{+0.08}_{-0.06}$ & $0.75^{+0.08}_{-0.1}$\\
\enddata
    \tablecomments{Values we consider unreliable are greyed out (see Section~\ref{sec_relations}).}
\end{deluxetable*}

%% file: ms.bbl
\begin{thebibliography}{}
\expandafter\ifx\csname natexlab\endcsname\relax\def\natexlab#1{#1}\fi
\providecommand{\url}[1]{\href{#1}{#1}}

\bibitem[{{Adams}(2010)}]{Adams2010}
{Adams}, F.~C. 2010, \araa, 48, 47

\bibitem[{{Adams} {et~al.}(2004){Adams}, {Hollenbach}, {Laughlin}, \&
  {Gorti}}]{Adams2004}
{Adams}, F.~C., {Hollenbach}, D., {Laughlin}, G., \& {Gorti}, U. 2004, \apj,
  611, 360

\bibitem[{{Alcal{\'a}} {et~al.}(2014){Alcal{\'a}}, {Natta}, {Manara}, {Spezzi},
  {Stelzer}, {Frasca}, {Biazzo}, {Covino}, {Randich}, {Rigliaco}, {Testi},
  {Comer{\'o}n}, {Cupani}, \& {D'Elia}}]{Alcala2014}
{Alcal{\'a}}, J.~M., {Natta}, A., {Manara}, C.~F., {et~al.} 2014, \aap, 561, A2

\bibitem[{{Allen} {et~al.}(2001){Allen}, {Bernstein}, \&
  {Malhotra}}]{Allen2001}
{Allen}, R.~L., {Bernstein}, G.~M., \& {Malhotra}, R. 2001, \apjl, 549, L241

\bibitem[{Anderson \& Darling(1952)}]{Anderson1952}
Anderson, T.~W., \& Darling, D.~A. 1952, The annals of mathematical statistics,
  23, 193

\bibitem[{{Andrews} {et~al.}(2013){Andrews}, {Rosenfeld}, {Kraus}, \&
  {Wilner}}]{Andrews2013}
{Andrews}, S.~M., {Rosenfeld}, K.~A., {Kraus}, A.~L., \& {Wilner}, D.~J. 2013,
  \apj, 771, 129

\bibitem[{{Andrews} {et~al.}(2018{\natexlab{a}}){Andrews}, {Terrell},
  {Tripathi}, {Ansdell}, {Williams}, \& {Wilner}}]{Andrews2018a}
{Andrews}, S.~M., {Terrell}, M., {Tripathi}, A., {et~al.} 2018{\natexlab{a}},
  \apj, 865, 157

\bibitem[{{Andrews} {et~al.}(2010){Andrews}, {Wilner}, {Hughes}, {Qi}, \&
  {Dullemond}}]{Andrews2010}
{Andrews}, S.~M., {Wilner}, D.~J., {Hughes}, A.~M., {Qi}, C., \& {Dullemond},
  C.~P. 2010, \apj, 723, 1241

\bibitem[{{Andrews} {et~al.}(2018{\natexlab{b}}){Andrews}, {Huang},
  {P{\'e}rez}, {Isella}, {Dullemond}, {Kurtovic}, {Guzm{\'a}n}, {Carpenter},
  {Wilner}, {Zhang}, {Zhu}, {Birnstiel}, {Bai}, {Benisty}, {Hughes},
  {{\"O}berg}, \& {Ricci}}]{Andrews2018}
{Andrews}, S.~M., {Huang}, J., {P{\'e}rez}, L.~M., {et~al.} 2018{\natexlab{b}},
  \apjl, 869, L41

\bibitem[{{Ansdell} {et~al.}(2016){Ansdell}, {Williams}, {van der Marel},
  {Carpenter}, {Guidi}, {Hogerheijde}, {Mathews}, {Manara}, {Miotello},
  {Natta}, {Oliveira}, {Tazzari}, {Testi}, {van Dishoeck}, \& {van
  Terwisga}}]{Ansdell2016}
{Ansdell}, M., {Williams}, J.~P., {van der Marel}, N., {et~al.} 2016, \apj,
  828, 46

\bibitem[{{Astropy Collaboration} {et~al.}(2013){Astropy Collaboration},
  {Robitaille}, {Tollerud}, {Greenfield}, {Droettboom}, {Bray}, {Aldcroft},
  {Davis}, {Ginsburg}, {Price-Whelan}, {Kerzendorf}, {Conley}, {Crighton},
  {Barbary}, {Muna}, {Ferguson}, {Grollier}, {Parikh}, {Nair}, {Unther},
  {Deil}, {Woillez}, {Conseil}, {Kramer}, {Turner}, {Singer}, {Fox}, {Weaver},
  {Zabalza}, {Edwards}, {Azalee Bostroem}, {Burke}, {Casey}, {Crawford},
  {Dencheva}, {Ely}, {Jenness}, {Labrie}, {Lim}, {Pierfederici}, {Pontzen},
  {Ptak}, {Refsdal}, {Servillat}, \& {Streicher}}]{AstropyCollaboration2013}
{Astropy Collaboration}, {Robitaille}, T.~P., {Tollerud}, E.~J., {et~al.} 2013,
  \aap, 558, A33

\bibitem[{{Astropy Collaboration} {et~al.}(2018){Astropy Collaboration},
  {Price-Whelan}, {Sip{\H{o}}cz}, {G{\"u}nther}, {Lim}, {Crawford}, {Conseil},
  {Shupe}, {Craig}, {Dencheva}, {Ginsburg}, {Vand erPlas}, {Bradley},
  {P{\'e}rez-Su{\'a}rez}, {de Val-Borro}, {Aldcroft}, {Cruz}, {Robitaille},
  {Tollerud}, {Ardelean}, {Babej}, {Bach}, {Bachetti}, {Bakanov}, {Bamford},
  {Barentsen}, {Barmby}, {Baumbach}, {Berry}, {Biscani}, {Boquien}, {Bostroem},
  {Bouma}, {Brammer}, {Bray}, {Breytenbach}, {Buddelmeijer}, {Burke},
  {Calderone}, {Cano Rodr{\'\i}guez}, {Cara}, {Cardoso}, {Cheedella}, {Copin},
  {Corrales}, {Crichton}, {D'Avella}, {Deil}, {Depagne}, {Dietrich}, {Donath},
  {Droettboom}, {Earl}, {Erben}, {Fabbro}, {Ferreira}, {Finethy}, {Fox},
  {Garrison}, {Gibbons}, {Goldstein}, {Gommers}, {Greco}, {Greenfield},
  {Groener}, {Grollier}, {Hagen}, {Hirst}, {Homeier}, {Horton}, {Hosseinzadeh},
  {Hu}, {Hunkeler}, {Ivezi{\'c}}, {Jain}, {Jenness}, {Kanarek}, {Kendrew},
  {Kern}, {Kerzendorf}, {Khvalko}, {King}, {Kirkby}, {Kulkarni}, {Kumar},
  {Lee}, {Lenz}, {Littlefair}, {Ma}, {Macleod}, {Mastropietro}, {McCully},
  {Montagnac}, {Morris}, {Mueller}, {Mumford}, {Muna}, {Murphy}, {Nelson},
  {Nguyen}, {Ninan}, {N{\"o}the}, {Ogaz}, {Oh}, {Parejko}, {Parley}, {Pascual},
  {Patil}, {Patil}, {Plunkett}, {Prochaska}, {Rastogi}, {Reddy Janga},
  {Sabater}, {Sakurikar}, {Seifert}, {Sherbert}, {Sherwood-Taylor}, {Shih},
  {Sick}, {Silbiger}, {Singanamalla}, {Singer}, {Sladen}, {Sooley},
  {Sornarajah}, {Streicher}, {Teuben}, {Thomas}, {Tremblay}, {Turner},
  {Terr{\'o}n}, {van Kerkwijk}, {de la Vega}, {Watkins}, {Weaver}, {Whitmore},
  {Woillez}, {Zabalza}, \& {Astropy Contributors}}]{AstropyCollaboration2018}
{Astropy Collaboration}, {Price-Whelan}, A.~M., {Sip{\H{o}}cz}, B.~M., {et~al.}
  2018, \aj, 156, 123

\bibitem[{{Bailer-Jones} {et~al.}(2018){Bailer-Jones}, {Rybizki}, {Fouesneau},
  {Mantelet}, \& {Andrae}}]{Bailer-Jones2018}
{Bailer-Jones}, C.~A.~L., {Rybizki}, J., {Fouesneau}, M., {Mantelet}, G., \&
  {Andrae}, R. 2018, \aj, 156, 58

\bibitem[{{Barenfeld} {et~al.}(2016){Barenfeld}, {Carpenter}, {Ricci}, \&
  {Isella}}]{Barenfeld2016}
{Barenfeld}, S.~A., {Carpenter}, J.~M., {Ricci}, L., \& {Isella}, A. 2016,
  \apj, 827, 142

\bibitem[{{Barenfeld} {et~al.}(2017){Barenfeld}, {Carpenter}, {Sargent},
  {Isella}, \& {Ricci}}]{Barenfeld2017}
{Barenfeld}, S.~A., {Carpenter}, J.~M., {Sargent}, A.~I., {Isella}, A., \&
  {Ricci}, L. 2017, \apj, 851, 85

\bibitem[{{Biazzo} {et~al.}(2017){Biazzo}, {Frasca}, {Alcal{\'a}}, {Zusi},
  {Covino}, {Randich}, {Esposito}, {Manara}, {Antoniucci}, {Nisini},
  {Rigliaco}, \& {Getman}}]{Biazzo2017}
{Biazzo}, K., {Frasca}, A., {Alcal{\'a}}, J.~M., {et~al.} 2017, \aap, 605, A66

\bibitem[{{Carpenter} {et~al.}(2014){Carpenter}, {Ricci}, \&
  {Isella}}]{Carpenter2014}
{Carpenter}, J.~M., {Ricci}, L., \& {Isella}, A. 2014, \apj, 787, 42

\bibitem[{{Chiang} \& {Goldreich}(1997)}]{Chiang1997}
{Chiang}, E.~I., \& {Goldreich}, P. 1997, \apj, 490, 368

\bibitem[{{Comer{\'o}n}(2008)}]{Comeron2008}
{Comer{\'o}n}, F. 2008, {The Lupus Clouds}, ed. B.~{Reipurth}, Vol.~5, 295

\bibitem[{{D'Alessio} {et~al.}(1998){D'Alessio}, {Cant{\"o}}, {Calvet}, \&
  {Lizano}}]{Dalessio1998}
{D'Alessio}, P., {Cant{\"o}}, J., {Calvet}, N., \& {Lizano}, S. 1998, \apj,
  500, 411

\bibitem[{Davidson-Pilon {et~al.}(2019)Davidson-Pilon, Kalderstam, Zivich,
  Kuhn, Fiore-Gartland, Moneda, Gabriel, WIlson, Parij, Stark, Anton, Besson,
  Jona, Gadgil, Golland, Hussey, Kumar, Noorbakhsh, Klintberg, Kaluzka,
  Slavitt, Martin, Ochoa, Albrecht, dhuynh, Zgonjanin, Chen, Fournier, Arturo,
  \& Rendeiro}]{cameron_davidson_pilon_2019_3240536}
Davidson-Pilon, C., Kalderstam, J., Zivich, P., {et~al.} 2019,
  CamDavidsonPilon/lifelines: v0.21.3, , , doi:10.5281/zenodo.3240536.
\newblock \url{https://doi.org/10.5281/zenodo.3240536}

\bibitem[{{de Zeeuw} {et~al.}(1999){de Zeeuw}, {Hoogerwerf}, {de Bruijne},
  {Brown}, \& {Blaauw}}]{deZeeuw1999}
{de Zeeuw}, P.~T., {Hoogerwerf}, R., {de Bruijne}, J.~H.~J., {Brown}, A.~G.~A.,
  \& {Blaauw}, A. 1999, \aj, 117, 354

\bibitem[{{Duncan} {et~al.}(1995){Duncan}, {Levison}, \& {Budd}}]{Duncan1995}
{Duncan}, M.~J., {Levison}, H.~F., \& {Budd}, S.~M. 1995, \aj, 110, 3073

\bibitem[{{Eisner} {et~al.}(2018){Eisner}, {Arce}, {Ballering}, {Bally},
  {Andrews}, {Boyden}, {Di Francesco}, {Fang}, {Johnstone}, {Kim}, {Mann},
  {Matthews}, {Pascucci}, {Ricci}, {Sheehan}, \& {Williams}}]{Eisner2018}
{Eisner}, J.~A., {Arce}, H.~G., {Ballering}, N.~P., {et~al.} 2018, \apj, 860,
  77

\bibitem[{{Evans}(1994)}]{Evans1994}
{Evans}, A. 1994, {The dusty universe}

\bibitem[{{Facchini} {et~al.}(2016){Facchini}, {Clarke}, \&
  {Bisbas}}]{Facchini2016}
{Facchini}, S., {Clarke}, C.~J., \& {Bisbas}, T.~G. 2016, \mnras, 457, 3593

\bibitem[{{Foreman-Mackey} {et~al.}(2013){Foreman-Mackey}, {Hogg}, {Lang}, \&
  {Goodman}}]{Foreman-Mackey2013}
{Foreman-Mackey}, D., {Hogg}, D.~W., {Lang}, D., \& {Goodman}, J. 2013, \pasp,
  125, 306

\bibitem[{{Frasca} {et~al.}(2017){Frasca}, {Biazzo}, {Alcal{\'a}}, {Manara},
  {Stelzer}, {Covino}, \& {Antoniucci}}]{Frasca2017}
{Frasca}, A., {Biazzo}, K., {Alcal{\'a}}, J.~M., {et~al.} 2017, \aap, 602, A33

\bibitem[{{Gerbig} {et~al.}(2019){Gerbig}, {Lenz}, \& {Klahr}}]{Gerbig2019}
{Gerbig}, K., {Lenz}, C.~T., \& {Klahr}, H. 2019, \aap, 629, A116

\bibitem[{{Guilloteau} {et~al.}(2011){Guilloteau}, {Dutrey}, {Pi{\'e}tu}, \&
  {Boehler}}]{Guilloteau2011}
{Guilloteau}, S., {Dutrey}, A., {Pi{\'e}tu}, V., \& {Boehler}, Y. 2011, \aap,
  529, A105

\bibitem[{{Huang} {et~al.}(2018){Huang}, {Andrews}, {Dullemond}, {Isella},
  {P{\'e}rez}, {Guzm{\'a}n}, {{\"O}berg}, {Zhu}, {Zhang}, {Bai}, {Benisty},
  {Birnstiel}, {Carpenter}, {Hughes}, {Ricci}, {Weaver}, \&
  {Wilner}}]{Huang2018}
{Huang}, J., {Andrews}, S.~M., {Dullemond}, C.~P., {et~al.} 2018, \apjl, 869,
  L42

\bibitem[{{Ida} {et~al.}(2000){Ida}, {Larwood}, \& {Burkert}}]{Ida2000}
{Ida}, S., {Larwood}, J., \& {Burkert}, A. 2000, \apj, 528, 351

\bibitem[{{Isella} {et~al.}(2009){Isella}, {Carpenter}, \&
  {Sargent}}]{Isella2009}
{Isella}, A., {Carpenter}, J.~M., \& {Sargent}, A.~I. 2009, \apj, 701, 260

\bibitem[{{Kelly}(2007)}]{Kelly2007}
{Kelly}, B.~C. 2007, \apj, 665, 1489

\bibitem[{{Kenyon} \& {Bromley}(2004)}]{Kenyon2004}
{Kenyon}, S.~J., \& {Bromley}, B.~C. 2004, \nat, 432, 598

\bibitem[{{Krijt} {et~al.}(2015){Krijt}, {Ormel}, {Dominik}, \&
  {Tielens}}]{Krijt2015}
{Krijt}, S., {Ormel}, C.~W., {Dominik}, C., \& {Tielens}, A.~G.~G.~M. 2015,
  \aap, 574, A83

\bibitem[{{Lauer} {et~al.}(1995){Lauer}, {Ajhar}, {Byun}, {Dressler}, {Faber},
  {Grillmair}, {Kormendy}, {Richstone}, \& {Tremaine}}]{Lauer1995}
{Lauer}, T.~R., {Ajhar}, E.~A., {Byun}, Y.~I., {et~al.} 1995, \aj, 110, 2622

\bibitem[{{Lenz} {et~al.}(2019){Lenz}, {Klahr}, \& {Birnstiel}}]{Lenz2019}
{Lenz}, C.~T., {Klahr}, H., \& {Birnstiel}, T. 2019, \apj, 874, 36

\bibitem[{{Long} {et~al.}(2018){Long}, {Herczeg}, {Pascucci}, {Apai},
  {Henning}, {Manara}, {Mulders}, {Sz{\H{u}}cs}, \& {Hendler}}]{Long2018}
{Long}, F., {Herczeg}, G.~J., {Pascucci}, I., {et~al.} 2018, \apj, 863, 61

\bibitem[{{Long} {et~al.}(2019){Long}, {Herczeg}, {Harsono}, {Pinilla},
  {Tazzari}, {Manara}, {Pascucci}, {Cabrit}, {Nisini}, {Johnstone}, {Edwards},
  {Salyk}, {Menard}, {Lodato}, {Boehler}, {Mace}, {Liu}, {Mulders}, {Hendler},
  {Ragusa}, {Fischer}, {Banzatti}, {Rigliaco}, {van de Plas}, {Dipierro},
  {Gully-Santiago}, \& {Lopez-Valdivia}}]{Long2019}
{Long}, F., {Herczeg}, G.~J., {Harsono}, D., {et~al.} 2019, \apj, 882, 49

\bibitem[{{Luhman}(2004)}]{Luhman2004}
{Luhman}, K.~L. 2004, \apj, 617, 1216

\bibitem[{{Luhman} \& {Rieke}(1999)}]{Luhman1999}
{Luhman}, K.~L., \& {Rieke}, G.~H. 1999, \apj, 525, 440

\bibitem[{{Luhman} {et~al.}(2008){Luhman}, {Allen}, {Allen}, {Gutermuth},
  {Hartmann}, {Mamajek}, {Megeath}, {Myers}, \& {Fazio}}]{Luhman2008}
{Luhman}, K.~L., {Allen}, L.~E., {Allen}, P.~R., {et~al.} 2008, \apj, 675, 1375

\bibitem[{{Manara} {et~al.}(2016){Manara}, {Fedele}, {Herczeg}, \&
  {Teixeira}}]{Manara2016}
{Manara}, C.~F., {Fedele}, D., {Herczeg}, G.~J., \& {Teixeira}, P.~S. 2016,
  \aap, 585, A136

\bibitem[{{Manara} {et~al.}(2017){Manara}, {Testi}, {Herczeg}, {Pascucci},
  {Alcal{\'a}}, {Natta}, {Antoniucci}, {Fedele}, {Mulders}, {Henning},
  {Mohanty}, {Prusti}, \& {Rigliaco}}]{Manara2017}
{Manara}, C.~F., {Testi}, L., {Herczeg}, G.~J., {et~al.} 2017, \aap, 604, A127

\bibitem[{{Manara} {et~al.}(2019){Manara}, {Tazzari}, {Long}, {Herczeg},
  {Lodato}, {Rota}, {Cazzoletti}, {van der Plas}, {Pinilla}, {Dipierro},
  {Edwards}, {Harsono}, {Johnstone}, {Liu}, {Menard}, {Nisini}, {Ragusa},
  {Boehler}, \& {Cabrit}}]{Manara2019}
{Manara}, C.~F., {Tazzari}, M., {Long}, F., {et~al.} 2019, \aap, 628, A95

\bibitem[{{McMullin} {et~al.}(2007){McMullin}, {Waters}, {Schiebel}, {Young},
  \& {Golap}}]{McMullin2007}
{McMullin}, J.~P., {Waters}, B., {Schiebel}, D., {Young}, W., \& {Golap}, K.
  2007, in Astronomical Society of the Pacific Conference Series, Vol. 376,
  Astronomical Data Analysis Software and Systems XVI, ed. R.~A. {Shaw},
  F.~{Hill}, \& D.~J. {Bell}, 127

\bibitem[{{Mer{\'\i}n} {et~al.}(2008){Mer{\'\i}n}, {J{\o}rgensen}, {Spezzi},
  {Alcal{\'a}}, {Evans}, {Harvey}, {Prusti}, {Chapman}, {Huard}, {van
  Dishoeck}, \& {Comer{\'o}n}}]{Merin2008}
{Mer{\'\i}n}, B., {J{\o}rgensen}, J., {Spezzi}, L., {et~al.} 2008, \apjs, 177,
  551

\bibitem[{{Morbidelli} {et~al.}(2008){Morbidelli}, {Levison}, \&
  {Gomes}}]{Morbidelli2008}
{Morbidelli}, A., {Levison}, H.~F., \& {Gomes}, R. 2008, {The Dynamical
  Structure of the Kuiper Belt and Its Primordial Origin}, ed. M.~A. {Barucci},
  H.~{Boehnhardt}, D.~P. {Cruikshank}, A.~{Morbidelli}, \& R.~{Dotson},
  275--292

\bibitem[{{Mulders} {et~al.}(2015){Mulders}, {Pascucci}, \&
  {Apai}}]{Mulders2015}
{Mulders}, G.~D., {Pascucci}, I., \& {Apai}, D. 2015, \apj, 798, 112

\bibitem[{{Nakajima} {et~al.}(2000){Nakajima}, {Tamura}, {Oasa}, \&
  {Nakajima}}]{Nakajima2000}
{Nakajima}, Y., {Tamura}, M., {Oasa}, Y., \& {Nakajima}, T. 2000, \aj, 119, 873

\bibitem[{{Ormel} {et~al.}(2017){Ormel}, {Liu}, \& {Schoonenberg}}]{Ormel2017}
{Ormel}, C.~W., {Liu}, B., \& {Schoonenberg}, D. 2017, \aap, 604, A1

\bibitem[{{Pascucci} {et~al.}(2018){Pascucci}, {Mulders}, {Gould}, \& {Fernand
  es}}]{Pascucci2018}
{Pascucci}, I., {Mulders}, G.~D., {Gould}, A., \& {Fernand es}, R. 2018, \apjl,
  856, L28

\bibitem[{{Pascucci} {et~al.}(2004){Pascucci}, {Wolf}, {Steinacker},
  {Dullemond}, {Henning}, {Niccolini}, {Woitke}, \& {Lopez}}]{Pascucci2004}
{Pascucci}, I., {Wolf}, S., {Steinacker}, J., {et~al.} 2004, \aap, 417, 793

\bibitem[{{Pascucci} {et~al.}(2016){Pascucci}, {Testi}, {Herczeg}, {Long},
  {Manara}, {Hendler}, {Mulders}, {Krijt}, {Ciesla}, {Henning}, {Mohanty},
  {Drabek-Maunder}, {Apai}, {Sz{\H{u}}cs}, {Sacco}, \&
  {Olofsson}}]{Pascucci2016}
{Pascucci}, I., {Testi}, L., {Herczeg}, G.~J., {et~al.} 2016, \apj, 831, 125

\bibitem[{{Pecaut} {et~al.}(2012){Pecaut}, {Mamajek}, \& {Bubar}}]{Pecaut2012}
{Pecaut}, M.~J., {Mamajek}, E.~E., \& {Bubar}, E.~J. 2012, \apj, 746, 154

\bibitem[{{Petit} {et~al.}(2011){Petit}, {Kavelaars}, {Gladman}, {Jones},
  {Parker}, {Van Laerhoven}, {Nicholson}, {Mars}, {Rousselot}, {Mousis},
  {Marsden}, {Bieryla}, {Taylor}, {Ashby}, {Benavidez}, {Campo Bagatin}, \&
  {Bernabeu}}]{Petit2011}
{Petit}, J.-M., {Kavelaars}, J.~J., {Gladman}, B.~J., {et~al.} 2011, \aj, 142,
  131

\bibitem[{{Pinilla} {et~al.}(2013){Pinilla}, {Birnstiel}, {Benisty}, {Ricci},
  {Natta}, {Dullemond}, {Dominik}, \& {Testi}}]{Pinilla2013}
{Pinilla}, P., {Birnstiel}, T., {Benisty}, M., {et~al.} 2013, \aap, 554, A95

\bibitem[{{Preibisch} {et~al.}(2002){Preibisch}, {Brown}, {Bridges},
  {Guenther}, \& {Zinnecker}}]{Preibisch2002}
{Preibisch}, T., {Brown}, A. G.~A., {Bridges}, T., {Guenther}, E., \&
  {Zinnecker}, H. 2002, \aj, 124, 404

\bibitem[{{Ricci} {et~al.}(2010){Ricci}, {Testi}, {Natta}, {Neri}, {Cabrit}, \&
  {Herczeg}}]{Ricci2010}
{Ricci}, L., {Testi}, L., {Natta}, A., {et~al.} 2010, \aap, 512, A15

\bibitem[{{Roccatagliata} {et~al.}(2018){Roccatagliata}, {Sacco},
  {Franciosini}, \& {Rand ich}}]{Roccatagliata2018}
{Roccatagliata}, V., {Sacco}, G.~G., {Franciosini}, E., \& {Rand ich}, S. 2018,
  \aap, 617, L4

\bibitem[{{Rosotti} {et~al.}(2019){Rosotti}, {Booth}, {Tazzari}, {Clarke},
  {Lodato}, \& {Testi}}]{Rosotti2019}
{Rosotti}, G.~P., {Booth}, R.~A., {Tazzari}, M., {et~al.} 2019, \mnras, 486,
  L63

\bibitem[{{Sacco} {et~al.}(2017){Sacco}, {Spina}, {Randich}, {Palla}, {Parker},
  {Jeffries}, {Jackson}, {Meyer}, {Mapelli}, {Lanzafame}, {Bonito}, {Damiani},
  {Franciosini}, {Frasca}, {Klutsch}, {Prisinzano}, {Tognelli},
  {Degl'Innocenti}, {Prada Moroni}, {Alfaro}, {Micela}, {Prusti}, {Barrado},
  {Biazzo}, {Bouy}, {Bravi}, {Lopez-Santiago}, {Wright}, {Bayo}, {Gilmore},
  {Bragaglia}, {Flaccomio}, {Koposov}, {Pancino}, {Casey}, {Costado}, {Donati},
  {Hourihane}, {Jofr{\'e}}, {Lardo}, {Lewis}, {Magrini}, {Monaco},
  {Morbidelli}, {Sousa}, {Worley}, \& {Zaggia}}]{Sacco2017}
{Sacco}, G.~G., {Spina}, L., {Randich}, S., {et~al.} 2017, \aap, 601, A97

\bibitem[{Shapiro \& Wilk(1965)}]{Shapiro1965}
Shapiro, S.~S., \& Wilk, M.~B. 1965, Biometrika, 52, 591.
\newblock \url{https://doi.org/10.1093/biomet/52.3-4.591}

\bibitem[{{Slesnick} {et~al.}(2008){Slesnick}, {Hillenbrand}, \&
  {Carpenter}}]{Slesnick2008}
{Slesnick}, C.~L., {Hillenbrand}, L.~A., \& {Carpenter}, J.~M. 2008, \apj, 688,
  377

\bibitem[{Tazzari(2017)}]{Tazzariuvplot}
Tazzari, M. 2017, doi:10.5281/zenodo.1003113

\bibitem[{{Tazzari} {et~al.}(2018){Tazzari}, {Beaujean}, \&
  {Testi}}]{Tazzari2018}
{Tazzari}, M., {Beaujean}, F., \& {Testi}, L. 2018, \mnras, 476, 4527

\bibitem[{{Tazzari} {et~al.}(2017){Tazzari}, {Testi}, {Natta}, {Ansdell},
  {Carpenter}, {Guidi}, {Hogerheijde}, {Manara}, {Miotello}, {van der Marel},
  {van Dishoeck}, \& {Williams}}]{Tazzari2017}
{Tazzari}, M., {Testi}, L., {Natta}, A., {et~al.} 2017, \aap, 606, A88

\bibitem[{{Tripathi} {et~al.}(2017){Tripathi}, {Andrews}, {Birnstiel}, \&
  {Wilner}}]{Tripathi2017}
{Tripathi}, A., {Andrews}, S.~M., {Birnstiel}, T., \& {Wilner}, D.~J. 2017,
  \apj, 845, 44

\bibitem[{{Wasserburg} {et~al.}(2017){Wasserburg}, {Karakas}, \&
  {Lugaro}}]{Wasserburg2017}
{Wasserburg}, G.~J., {Karakas}, A.~I., \& {Lugaro}, M. 2017, \apj, 836, 126

\bibitem[{{Wilking} {et~al.}(2005){Wilking}, {Meyer}, {Robinson}, \&
  {Greene}}]{Wilking2005}
{Wilking}, B.~A., {Meyer}, M.~R., {Robinson}, J.~G., \& {Greene}, T.~P. 2005,
  \aj, 130, 1733

\bibitem[{{Winter} {et~al.}(2018){Winter}, {Clarke}, {Rosotti}, {Ih},
  {Facchini}, \& {Haworth}}]{Winter2018}
{Winter}, A.~J., {Clarke}, C.~J., {Rosotti}, G., {et~al.} 2018, \mnras, 478,
  2700

\bibitem[{{Zinner}(2014)}]{Zinner2014}
{Zinner}, E. 2014, {Presolar Grains}, ed. A.~M. {Davis}, Vol.~1, 181--213

\end{thebibliography}
